\documentclass{aastex}
\usepackage{emulateapj5}
\newif\ifAMStwofonts
\AMStwofontstrue

\def\hi{{\mbox{\sc H i}}}

\def\kms{km~s$^{-1}$}
\def\apss{Ap\&SS}
\def\degree{$^{\circ}$}

\def\ga{\mathrel{\hbox{\rlap{\hbox{\lower4pt\hbox{$\sim$}}}\hbox{$>$}}}}
\def\la{\mathrel{\hbox{\rlap{\hbox{\lower4pt\hbox{$\sim$}}}\hbox{$<$}}}}

\shorttitle{Pulsar Measurements of  Tiny-Scale Atomic Structure}

\shortauthors{S.\ Stanimirovi\'{c} et al.}
\submitted{Accepted by ApJ}

\begin{document}

\title{Arecibo Multi-Epoch HI Absorption Measurements Against Pulsars: Tiny-Scale Atomic Structure}

\author{S. Stanimirovi\'{c}$^{1}$, J. M. Weisberg$^{2}$, Z. Pei$^{2}$, K. Tuttle$^{2}$,
J. T. Green$^{2}$}
\setcounter{footnote}{1}
\footnotetext{Department of Astronomy, University of Wisconsin, Madison, WI 53706; sstanimi@astro.wisc.edu}
\setcounter{footnote}{2}
\footnotetext{Department of Physics and Astronomy, Carleton College,
Northfield, MN 55057}

\begin{abstract}
We present results from multi-epoch neutral hydrogen (HI) absorption observations
of six bright pulsars with the Arecibo telescope.
Moving through the interstellar medium (ISM) with  transverse velocities of 10--150 AU yr$^{-1}$, 
these pulsars have swept across 1--200 AU over the course of our experiment, allowing us to 
probe the existence and properties of the tiny scale atomic structure (TSAS) in the cold neutral medium (CNM).
While most of the observed pulsars show no significant change in their HI absorption spectra,
we have identified at least two clear TSAS-induced opacity variations
in the direction of B1929+10. These observations require strong spatial inhomogeneities
in either the TSAS clouds' physical properties themselves or else in the clouds'  galactic distribution.  
While TSAS is occasionally detected on spatial scales down to 10 AU, it is too rare to be
characterized by a spectrum of turbulent CNM fluctuations  on scales of $10^{1-3}$
AU, as previously suggested by some work. In the direction of B1929+10, an
apparent correlation between TSAS and interstellar clouds inside the warm Local Bubble (LB) indicates
that TSAS may be tracing the fragmentation
of the LB wall via hydrodynamic instabilities.
While similar fragmentation events occur frequently throughout the ISM,
the warm medium surrounding these cold cloudlets  induces  a natural selection effect
wherein small TSAS clouds evaporate quickly and are rare,
while large clouds survive longer and become a general property
of the ISM.
\end{abstract}

\keywords{ISM: clouds --- ISM: structure --- line: profiles}

\section{Introduction}

For many years, observations of the diffuse interstellar medium (ISM) have traced a whole 
hierarchy of structures on spatial scales $\ga1$ pc (cf. \cite{Dickey90}), often attributed to the ISM turbulence.
However, the extremely small-scale end of this
spectrum, reaching down to scales of tens to hundreds of AUs, has been largely unexplored. 
As the kinetic energy cascades from larger to smaller scales, it is expected
that the turbulent spectrum will reach its end at so-called ``dissipation scales.''
Understanding of the dissipative properties of turbulence, such as its scale, efficiency, 
dissipation rate, and ubiquity, are  important since
the delicate balance between energy injection and dissipation
has a profound impact on the stellar -- ISM recycling chain.
However, many questions still await both theoretical understanding and
observational constraints.
For example, how is the balance between turbulent energy injection and dissipation maintained?
What is the efficiency of the dissipation process? Does it vary
with time and galactocentric radius? Is dissipation located only in pockets
of ``active turbulence'', or it is ubiquitous?
How do dissipative properties vary with galactic environments?

Starting in 1976  \citep{Dieter76}, structure in the diffuse ISM has been observed 
sporadically on scales  down to $10^{1-2}$ AU.  These findings caused much controversy as the
observationally inferred properties were not in accord with the standard ISM picture.
With a typical, observationally inferred, HI volume density of $\sim10^{4}$ cm$^{-3}$,
and the thermal pressure of $\sim10^{6}$ cm$^{-3}$ K (assuming
temperature of $\sim$40--70 K), the
AU-scale ``tiny'' features appeared significantly over-dense and over-pressured compared with 
the traditional cold neutral medium (CNM) clouds.

Since the initial observations, the AU-sized structures came to be rather frequently detected, with claims that
up to 15\% of the CNM could be in this form \citep{Frail94}. There has been considerable progress in recent years in
understanding of the ISM turbulence and the small-scale structure in the ISM, from both 
theoretical and observational perspectives [for reviews see \cite {Elmegreen04,Scalo04}]. Nevertheless, 
it remains unclear
whether the observed AU-scale features represent the dissipation scales of the $\ga1$ pc spatial scale turbulence.
At the same time, the observational picture has become even more complicated with
improved telescope sensitivity, resulting in the detection
of new populations of possibly related sub-parsec ISM clouds
\citep{Braun05, Stanimirovic05, Begum10}.
The ``low column density clouds'' observed by Stanimirovic \& Heiles (2005)
have a peak optical depth of only $10^{-3}$ to $10^{-2}$ and an inferred size of 500--5000 AU.

Motivated by the recent theoretical and observational efforts in understanding the 
nature and origin of the tiny-scale ISM structure, we have undertaken multi-epoch 
observations of the HI absorption against a set of bright pulsars. Because of pulsars' 
relatively high proper motion and transverse speeds
(typically $\sim10 - 10^{2}$ AU yr$^{-1}$), pulsar HI absorption profiles
obtained at different epochs sample CNM structure on AU spatial scales.
In addition, the pulsed nature of pulsars' emission allows spectra to
be obtained {\it on} and {\it off} source without moving the
telescope, therefore sampling both emission and absorption along almost exactly the same line of sight.
\cite{Stanimirovic03b} summarized preliminary results from this project for three 
pulsars, B0823+26, B1133+16 and B2016+28. Here we present results for all pulsars with full analysis.

The structure of this paper is organized in the following way.
Section~\ref{s:background} provides a short summary of the main observational 
and theoretical results of the AU-scale structure in neutral gas. In Section~\ref{s:obs} 
we summarize our observing and data processing strategies. Results from our investigation of 
changes in absorption profiles' {\it integrated} optical depth properties
(c.f. equivalent widths) are presented in  Section~\ref{s:EW}, while in Section~\ref{s:results} we search for
HI absorption {\it feature} optical depth variability.
To facilitate further data analysis we derive the
spin temperature and HI column density in the direction of all pulsars in
Section~\ref{s:gaussian_results}.
In Section 7, we investigate in detail the properties of the ISM along lines of sight toward our pulsars. 
In Section~\ref{s:sig_features} we study the properties of
the most significant AU-scale features found in this study.
We discuss our results and compare them with several theoretical models in
Section~\ref{s:discussion}, and summarize our findings in Section~\ref{s:conclusions}.

\section{Background}
\label{s:background}

While
structure on AU scales has been observed in neutral, ionized, and molecular
flavors of diffuse gas, it is still not clear whether all these features trace a single 
or multiple phenomena. Our work focuses only on the tiny-scale {\it{atomic}}
structure (TSAS; Heiles 1997) and we summarize results from two
different observational approaches used to study TSAS in the CNM below in
\S \ref{s:obsTSAS-VLBI} and \S \ref{s:obsTSAS-PSR}, as well as several theoretical 
considerations of the TSAS phenomenon in \S \ref{s:theoryTSAS}.
In addition to observational methods discussed here, TSAS in the CNM has been
also observed through temporal (probing scales of $\sim1$ to tens of AUs) and spatial
(probing scales of a few thousands of AU) variability of optical absorption 
lines \citep{Meyer96,Andrews01,Crawford00,Lauroesch03}.
The reader is referred to excellent articles in the ASP Conf. Series Vol. 365 \citep{Haverkorn07} 
for complementary details  on optical absorption line studies, or the Galactic 
tiny-scale structure in the  ionized and molecular gas. 
Finally, there are also some indications that even damped Lyman $\alpha$ absorbing systems 
at high redshift may contain small-scale HI structure on scales of $\sim10-100$ AU 
(Kanekar \& Chengalur 2001).

\subsection{Observations of TSAS via spatial mapping  of HI absorption line
profiles across  extended background sources}

\label{s:obsTSAS-VLBI}

Single baseline very long baseline interferometric
(VLBI) observations of the quasar 3C147 by \cite{Dieter76} were the first to infer the existence 
of a tiny-scale atomic cloud,  having a size of 70 AU and HI volume density of $10^{5}$ cm$^{-3}$, 
based on variations in absorption line profiles in the
directions of separate source components.
Subsequent VLBI observations by \cite{Diamond89}
found evidence for 25-AU HI clouds in directions toward 3C138, 3C380, and 3C147.
Particularly large variations were noticed in the case of
3C138.

The first {\it{images}} of HI optical depth distribution in the direction of
extragalactic sources were obtained by
\cite{Davis96} toward 3C138 and 3C147, using the MERLIN array and the European VLBI Network.
\cite{Faison98} and \cite{Faison01} used the Very Long Baseline Array (VLBA) to image 
HI absorption toward seven sources.
Significant variations in HI optical depth with $\Delta \tau$ of 0.1 and
$\sim1$, respectively, were found toward only two sources, 3C147 and 3C138.

The latest results on  3C138  by \cite{Brogan05}, in a three-epoch series of observations, 
show clear evidence for spatial variations of
$\Delta\tau\sim0.5$ on scales of 25 AU.
In addition, {\it{temporal}} changes in HI optical depth have been found over a
period of 7 years, with implied transverse
velocities of order 20 \kms~
(see \cite{Brogan07}  for a thorough review of these measurements). Some 10\% of 
pixels in their optical depth images have $\Delta
\tau >5$-$\sigma$. Assuming that most of the CNM along the line of sight
contains small-scale structure, they estimated the volume filling factor of about 1\%.
They proposed that TSAS is ubiquitous, and that the lower measured levels of variations 
in the other  cases result from  a selection effect due to incomplete sampling of the relevant 
angular and spatial scales.

Recently, \cite{Goss08} used MERLIN observations of  3C161
and found $\Delta \tau \sim $1.0-1.5 on scales of $\sim500$ AU.
A slightly smaller level of
variations (typically $\Delta \tau\sim0.2$)
was found in the case of 3C147, on scales of about 10 AU
\citep{Lazio09}.

\subsection{Observations of TSAS via time variability of HI absorption profiles against pulsars}

\label{s:obsTSAS-PSR}

In the late 1980s, sufficiently accurate and repeated measurements of HI absorption profiles
against pulsars began to be made, and it was
noticed that pulsar ISM spectra changed over time in some cases,
suggesting inhomogeneities in the intervening gas.  For
example, \citet{Clifton88} found that  the HI absorption spectrum of PSR B1821+05
changed significantly between ~1981 and 1988, with the appearance at the
latter epoch of a previously unobserved feature with $\tau\sim2$ and $\Delta v
\sim 1$ \kms.  \citet{Deshpande92} showed that between $\sim1976$ and 1981, HI
absorption toward  B1154-62 did  {\it{not}} change significantly; while toward
B1557-50, a variation with $\Delta\tau\sim1$ was interpreted as a cloud of
size in the 1000 AU range.

The early pulsar HI results inspired \cite{Frail94}
to undertake a dedicated multi-epoch pulsar HI absorption experiment at
Arecibo.  Six pulsars were observed at three epochs, with time baselines
ranging from 0.7--1.7 yr.  These authors reported the presence of
{\it{pervasive}} variations with $\Delta\tau\sim0.03-0.7$,  and associated HI
column densities of $10^{19}$ -- $5 \times 10^{20}$ cm$^{-2}$; on scales of 5--100
AU.  They indicated that 10-15\% of cold HI is in the tiny structures, and
additionally detected a correlation between absorption equivalent width
variations and equivalent width itself.  These results appeared to buttress
the earlier VLBI findings of TSAS, and they provided a strong
impetus for further experimental and theoretical work.

The recent era of  pulsar TSAS experiments began with the Parkes observations
of  \citet{Johnston03}.  Surprisingly, these investigators found no significant
optical depth variations in their three-epoch, 2.5-yr  observations of three
pulsars, and were able to place an upper limit on column density variations
toward PSR B1641-45 of $10^{19}$ cm$^{-2}$, significantly below the
\citet{Frail94} detections.  They showed that the earlier experiment did not fully
account for the large increases in noise in   absorption spectra at the line
frequency, so that some of the apparently significant  variations actually
were not.  While no significant variations were seen by \citet{Johnston03} during the
2.5 years of their TSAS experiment, they detected variations in the spectrum
of PSR B1557-50 when compared with measurements made five years earlier.  This
is the same pulsar whose spectrum was noted earlier by \citet{Deshpande92} to vary on
similar timescales in the late 1970s.  In combining the results from four
measurements over twenty-five years, \citet{Johnston03} concluded that the cloud
causing the variations is $\sim 1000$ AU in size, with a density of $\sim
10^4$ cm$^{-3}$.

\citet{Minter05} performed a particularly exhaustive TSAS study on PSR B0329+54,
which is very bright and almost circumpolar at the Green Bank Telescope. The
investigators observed the pulsar continuously for up to 20 hr in eighteen
observing sessions over a period of  1.3 yr.   They detected no HI optical depth
variations ($\Delta\tau < 0.026$ in most cases) for pulsar transverse offsets
ranging from 0.005--25 AU.

We and our colleagues published early results from a new set of multi-epoch Arecibo observations
\citep{Stanimirovic03b}. Based on non-detections of TSAS in
the direction of three pulsars (B0823+26, B1133+16, and B2016+28), we suggested that TSAS
is not ubiquitous in the ISM and could be related to isolated events.
The current work elaborates these themes.

\subsection{Theoretical considerations of TSAS}

\label{s:theoryTSAS}

Several explanations were proposed to reconcile the observations of TSAS with 
theory. \cite{Heiles97} proposed that TSAS features could be curved filaments and/or 
sheets that happen to be aligned along our line-of-sight.
By combining low temperature and a geometrical elongation, the apparently problematic 
volume density and thermal pressure can be significantly reduced.
We discuss this model in some detail in Section~\ref{s:heiles}.

\cite{Deshpande00a} suggested that TSAS represents the tail-end of the turbulent
ISM spectrum which exists on larger spatial scales, and showed that the observed optical 
depth differences are consistent with
a single power-law description of the HI optical depth distribution as a function of spatial 
scale. It was also pointed out that the over-density of TSAS is due to
a misinterpretation of observations: the observed variations in optical depth sample
the square-root of the structure function of the HI optical depth, not directly the power 
spectrum of optical depth fluctuations.
If the power spectrum of the HI optical depth distribution in the direction of
Cas A, with a measured slope of 2.75 \citep{Deshpande00b} over a range of spatial scales
from 0.02 to 4 pc, is extrapolated to AU scales then $\Delta \tau=0.2$-0.4 is expected on scales of 50-100 AU.
We note that this slope was confirmed recently
in an independent experiment by \cite{Roy10}.

\cite{Gwinn01} proposed that optical depth fluctuations
seen in multi-epoch pulsar observations are a scintillation
phenomenon combined with a velocity gradient across the absorbing HI.
While in the case of \cite{Deshpande00a} optical depth variations are expected
to increase with the size of structure, the scintillation phenomenon predicts maximum
variations on the very small spatial scales probed by interstellar scintillation. However, 
dedicated observations of PSR J1456-6843 by \cite{Gwinn07} failed to detect any scintillation 
effects on the HI absorption line.

One of the main problems regarding TSAS is its over-pressure relative to the
traditional thermal pressure of the ISM, which suggests that TSAS can not persist nor be pervasive.  However,
recent numerical simulations of the dynamic and turbulent ISM find substantial departures from 
pressure equilibrium. Instead of well-defined thermal phases, gas with a wide range of densities 
and temperatures is found in simulations \citep{MacLow05, Hennebelle07a}. Such results are
 supported by observations of the CI fine-structure lines by \cite{Jenkins01} 
who found evidence for the existence of
over-pressured gas ($P/k \ga 10^{5}$ cm$^{-3}$ K).
While these studies
suggest that high-pressure TSAS could be a natural product of
the highly dynamic ISM, the fraction of high-pressure gas in the ISM, as well as the 
physical processes responsible for their formation, vary hugely across
simulations. The only way to constrain TSAS properties and abundance is through sensitive observations.

\section{Observations and Data Processing}
\label{s:obs}

\begin{figure*}
\plotone{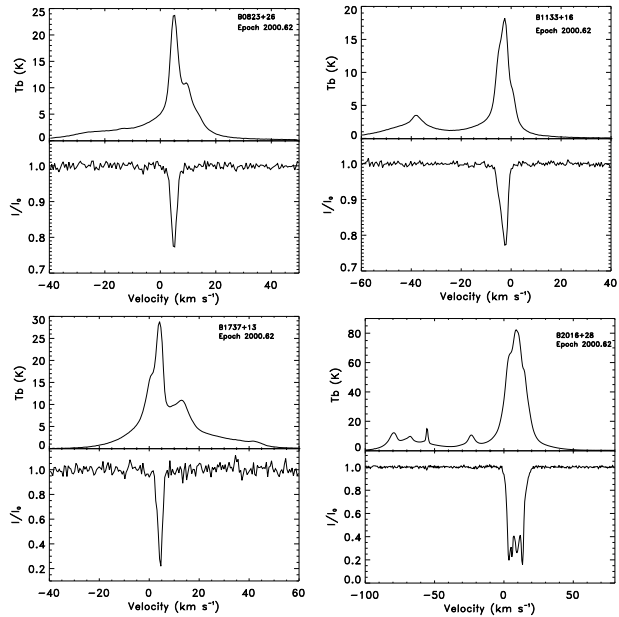}
\caption{\label{f:all_abs}Emission (top) and absorption (bottom) profiles in
the direction of B0823+26, B1133+16, B1737+13, and B2016+28
obtained at the first epoch (August 2000).
The emission
spectra were scaled to match the brightness temperature from the Leiden-Dwingeloo survey.}
\end{figure*}

\begin{figure*}
\plotone{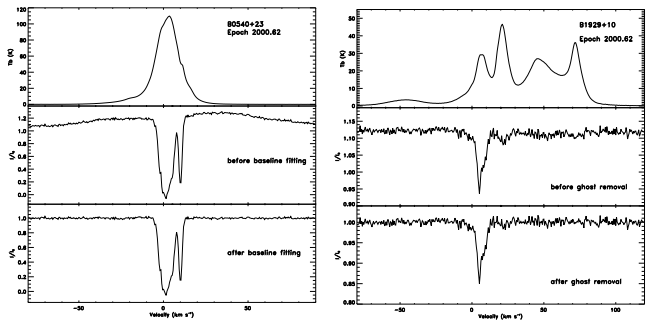}
\caption{\label{f:0540_1929em}Emission and absorption profiles in
the direction of B0540+23 and B1929+10, obtained at the first epoch (August 2000).
The middle panels show the absorption profiles before baseline fitting for B0540+23, 
and before ghost removal (see \S \ref{s:baseline}) for B1929+10.
In the case of B0540+23, a large ripple is obvious in the middle panel, caused by the
 interstellar scintillation. For B1929+10, a faint inverted copy of the \hi\ emission spectrum is noticeable
especially around 20 \kms.
The bottom panels show the absorption spectra after
baseline and ghost correction, respectively.}
\end{figure*}

\subsection{The observing and basic data processing procedure}

\begin{table*} \caption{Basic parameters of pulsars observed in this study, tabulated or 
derived using the reference given in the last column and the ATNF Pulsar Catalogue \citep{Manchester05}.}
\centering
\label{t:observed_psrs}
\begin{tabular}{ccccccccc}
\noalign{\smallskip} \hline \hline \noalign{\smallskip}
PSR   & $l/b$ & DM  & S$_{\rm 20}$ & D & V$_{\rm t}$& L$_{\perp}$& $T_{\rm PSR}$ & Reference \\
(B1950) & deg  &(pc\,cm$^{-3}$)& (mJy)& (kpc) & (\kms) & (AU yr$^{-1}$)& (K) & \\
\hline
B0540+23 & 184.36/$-3.32$ &77.7 & 9  & 3.5 & 377  & 80 &2.7 & \cite{Harrison93}\\
B0823+26 & 196.96/31.74   &19.5 & 10 & 0.4   & 192 & 39 &6.9& \cite{Gwinn86}\\
B1133+16 & 241.90/69.19   &4.8 & 32 &  0.4 & 631 & 135 &9.0& \cite{Brisken02}\\
B1737+13 & 37.08/21.68    &48.9 & 3.9 & 4.7 & 672 & 142 &0.7& \cite{Brisken03}\\
B1929+10 & 47.38/$-3.88$  &3.2  & 36 & 0.3 & 177  & 37 & 8.3 &\cite{Chatterjee04}\\
B2016+28 & 68.10/$-3.98$  &14.2 & 30 & 1.0 & 30 & 7 &8.4& \cite{Brisken02}\\
\noalign{\smallskip} \hline \noalign{\smallskip}
\end{tabular}
\end{table*}

We used the Arecibo telescope\footnote{The Arecibo Observatory
is part of the National Astronomy
and Ionosphere Center, operated by Cornell University under a
cooperative agreement with the National Science Foundation.}
and the Caltech Baseband Recorder (CBR) backend \citep{Jenet97}, to 
measure HI  absorption spectra of six pulsars at multiple epochs. We chose to observe the same
sources as Frail et al. (1994) in order to enhance the number of available
time baselines for comparison. Table~\ref{t:observed_psrs} lists several basic
parameters for our targets: Galactic coordinates ($l/b$), the dispersion measure (DM),
continuum equivalent flux density at 20
cm (S$_{20}$), distance (D), the transverse velocity (V$_{t}$), the transverse distance traveled 
in a year (L$_{\perp}$), and the pulsar-on antenna temperature ($T_{\rm PSR}$). 
The tabulated  flux density and $T_{\rm PSR}$ 
represent average values obtained from long integrations,
whereas the actual pulsar flux density and $T_{\rm PSR}$ vary on 
timescales from several minutes to several hours due to interstellar scintillation.
Distances for all six pulsars (column 5) were determined from
interferometric parallax measurements.
The pulsar-on antenna temperature $T_{\rm PSR}= G S_{20} P/W_{50}$, 
with the L-band gain $G=7.5 \times 10^{-3}$ K mJy$^{-1}$ (Heiles et al. 2001), and
pulsar period $P$ and 50\% pulse width $W_{50}$ from the ATNF pulsar Catalogue (Manchester et al. 2005).

The observing and basic data processing procedures were described in 
\cite{Stanimirovic03b} and \cite{Weisberg08}.
Relative to the previous large multi-epoch experiment by \cite{Frail94},
our four-level spectrometer is less prone to systematic errors when observing these strong 
and highly-variable sources. We had four observing sessions, referred to below as S1, S2, S3, S4; for 
epochs 2000.62, 2000.95, 2001.70, and 2001.86, respectively.
The typical integration time per pulsar was about 8 hours for each of four sessions.

The raw complex voltage samples recorded with the CBR were processed at Caltech's 
Center for Advanced Computation and Research to obtain a data cube of pulsar intensity as a function of 
pulsar rotational phase and radio frequency. For each scan, portions of the data 
cube during and off the pulsar pulse were accumulated
 to obtain the
`pulsar-on' and `pulsar-off' spectra. The pulsar absorption spectrum represents the `pulsar-on' --
`pulsar-off' spectrum for each scan, whose baseline was then flattened by performing frequency 
switching and normalizing by the off--line pulsar intensity.
In accumulating pulsar absorption spectra from different $\sim 2-$min scans, each spectrum
was given a weight proportional to $T^2_{\rm PSR}$, where $T_{\rm PSR}$ is the antenna temperature 
of the pulsar.

The frequency-switched `pulsar-off' spectra correspond to the \hi\ emission spectra
in pulsar directions. As no temperature calibration was performed during the observations, we have 
scaled these spectra to match the \hi\ brightness temperature from the 
Leiden-Dwingeloo survey \citep{Burton94}.
Final absorption and emission spectra for all pulsars, from the first observing epoch (2000.62), are shown in 
Figures~\ref{f:all_abs} and~\ref{f:0540_1929em}.
Our spectrometer, consisting of  4096 frequency channels across 10 MHz total bandwidth,  
has a velocity channel spacing of 0.52-\kms.  
However, adjacent spectral channels are not independent, which can lead to
distortion of narrowband signals such as the ones we are studying \citep{Tools}. 
To ameliorate this problem, we
Hanning smoothed all processed absorption spectra, yielding a final
velocity resolution of 1.04 \kms.
Note that Hanning smoothing was not applied in our preliminary analyses [Stanimirovic et al. (2003); 
Weisberg \& Stanimirovic (2007)].

\subsection{Baseline fitting of absorption spectra}
\label{s:baseline}

To flatten the baseline of the final absorption spectra
we fitted a polynomial function to the unabsorbed portions of the  spectrum.
For pulsars B0823+26, B1133+16, B1737+13 and B2016+28 generally a 1st or 2nd order polynomial
function was fitted and then the full spectrum was divided by the fitted function. A higher order 
polynomial was required for the 3rd session data for B2016+28 due to ripples in the baseline 
structure, most likely cased by interstellar scintillation.

PSR B0540+23 data were heavily affected by interstellar
scintillation. This pulsar's scintillation bandwidth at 1.4 GHz is 0.3 MHz \citep{Gwinn01},
which resulted in numerous ripples across our band (see
Figure~\ref{f:0540_1929em}, left middle panel). We needed to fit and divide by a 15th order 
polynomial to flatten the baseline for this pulsar (see bottom panel in Figure~\ref{f:0540_1929em}).
However, as emphasized in Section 4.1, we do not use  this pulsar's data for
our analysis due to large scintillation effects which presumably led to the unphysically ($I/I_{0} < 0$) deep
absorption features.
Noise and calibration errors in the pulsar absorption spectrum are magnified
by a factor of  $e^{\tau}$ when the absorption spectrum is expressed in terms of
the  ultimately desired optical depth $\tau$. With this pulsar's optical depth near infinity, 
these errors become so large as to render our measurements unusable.

B1929+10 data, as shown in Figure~\ref{f:0540_1929em} (right, middle panel), suffer from 
the `ghost effect' \citep{Weisberg80}.
The main absorption feature is at 5 \kms. However several additional
features are visible from 20 to 80 \kms~which represent
a false image of the pulsar--on emission spectrum. The usual reason for the
ghosts is an incorrect scaling of total power received by the
spectrometer in the case of strong and variable sources. As the ghost is a faint copy of 
the pulsar--on emission spectrum in the pulsar absorption spectrum,
we can exorcize the ghost by fitting for and removing the former
from the latter.

To do so we have performed a
least-squares fit of the $I(v)/I_{0}$ absorption spectrum to a ghost of fitted amplitude $A$, and a 
fitted linear baseline slope $B$ plus constant $C$, all
to spectral channels lacking true absorption: \begin{equation}
I(v)/I_{0} = A T_{\rm PSR-on}(v) + Bv + C,
\end{equation}
where $T_{\rm PSR-on}(v)$ is the measured pulsar-on spectrum. The exorcized
absorption spectrum is then: \begin{equation}
(I(v)/I_{0})_{\rm corrected} = [I(v)/I_{0}-A T_{\rm PSR-on}(v)-Bv]/C.
\end{equation}

The exorcized PSR B1929+10 absorption spectrum is shown in Figure~\ref{f:0540_1929em}, 
right bottom panel. The fitted parameters are listed in Table~\ref{t:ghost}.

\begin{table*} \caption{
Fitted parameters for the `ghost' excision in the case of B1929+10
(Section 3.2).}
\centering
\label{t:ghost}
\begin{tabular}{cccc}
\noalign{\smallskip} \hline \hline \noalign{\smallskip}
Session   & $A$ & $B$  & $C$ \\
 &(K$^{-1}$) & ($10^{-5}$ km$^{-1}$ s)&   \\
\hline
1 &$-4.3\pm0.4$  &$-3.0\pm0.6$ & $1.148\pm0.003$\\
2 & $-9.6\pm0.4$  &$6.1\pm0.3$ & $1.147\pm0.002$\\
3 & $-7.5\pm0.4$  &$4.8\pm0.3$ & $1.128\pm0.002$\\
4 & $-14.9\pm0.7$   &$-0.3\pm 0.3$ & $1.144\pm0.002$\\
\noalign{\smallskip} \hline \noalign{\smallskip}
\end{tabular}
\end{table*}

\subsection{Noise envelopes}
\label{s:noise}

For each absorption spectrum we have calculated an expected noise envelope
which will be used to assess the significance of any observed temporal variations.
The expected  noise level $\Delta T(v)$ is proportional to the overall system temperature $T_{\rm sys}(v)$,
which has the following contributions.
(1) The receiver temperature $T_{\rm rx}\sim25$--30 K.
(2) The sky background, which is typically of order of
1--2 K, based on the all-sky survey at 408 MHz by \cite{Haslam82} and after scaling 
to 1.4 GHz using a spectral index of $-2.6$. (3) The contribution from the HI {\em {emission}} which
can be very significant and which varies with velocity. 
For example the rms noise on--line is three times higher
than the rms noise off--line for the case of B0540+23.
(4) At Arecibo, the  pulsar continuum emission itself is also  sometimes sufficiently 
strong to provide the principal contribution to the system temperature. 

Since an observed pulsar-on spectrum is identically the sum of these four contributions, 
the expected noise in the pulsar-on spectrum 
should be proportional to the pulsar-on spectrum.
Furthermore, owing to pulsar-off integration times being $\sim10-100$ times longer than
pulsar-on integration times, the pulsar-off spectrum adds negligible additional
noise to the absorption spectrum. 
 Hence our expected noise envelopes (shown in 
Figures~\ref{f:0823_6plot}--\ref{f:2016_6plot})
are generated by scaling the observed pulsar-on spectra
by a multiplicative factor that matches them to the observed noise 
in the off-line portions of the absorption spectra.
We have then propagated uncertainties associated
with baseline fitting and ghost removal procedures to derive the final expected noise envelopes.
To estimate significance envelopes for
{\it{difference}} spectra shown in Section~\ref{s:results} (Figures 3-8)
or for the equivalent width analysis in Section~\ref{s:EW},
appropriate noise envelopes for individual absorption profiles were added in quadrature.

To investigate the properties of noise in emission-free channels,
we have plotted the histogram of data points located off-line for each difference spectrum 
(not shown in the paper). Most histograms are consistent with Gaussian statistics with a 
standard deviation in agreement or close to our derived 1-$\sigma$ noise level.
In several cases there is a small discrepancy between the measured off-line
noise and derived 1-$\sigma$ noise envelopes. This is not surprising after the 
propagation of errors due to baseline, and especially ghost, fitting.
For at least three of our pulsars we have achieved excellent sensitivity.
We discuss the rms on-line noise level of the final absorption spectra in Section~\ref{s:results}.

\section{Equivalent width variations}
\label{s:EW}

Our main goal is to search for variations in the HI absorption profiles
obtained at different observing epochs.
We have employed two methods: (i)  in this section we look for
changes in integrated properties such as the equivalent width, and
(ii)  in Section~\ref{s:results} we investigate spectra differenced between
each pair of observing epochs.

\begin{figure*}
\epsscale{1.5}
\plotone{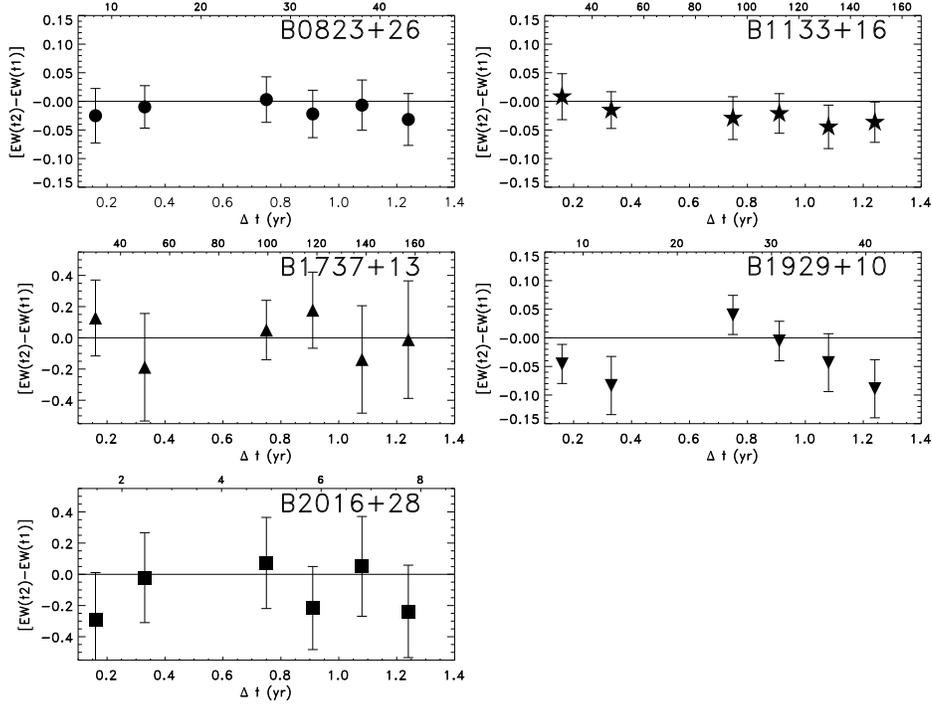}
\vspace{0.5cm}
\caption{\label{f:comparison_EW}
Temporal change in the equivalent width between two epochs,
$\Delta {\rm EW}(\Delta t) = [{\rm EW}(t_{2}) - {\rm
 EW}(t_{1})]$. Plotted error bars 
represent $\pm2$-$\sigma$ uncertainties. The top x-axis on
 each plot
 shows the corresponding spatial scales, $l=\Delta t \times L_{\perp}$.}
\end{figure*}

\begin{figure*}
\epsscale{1.2}
\plotone{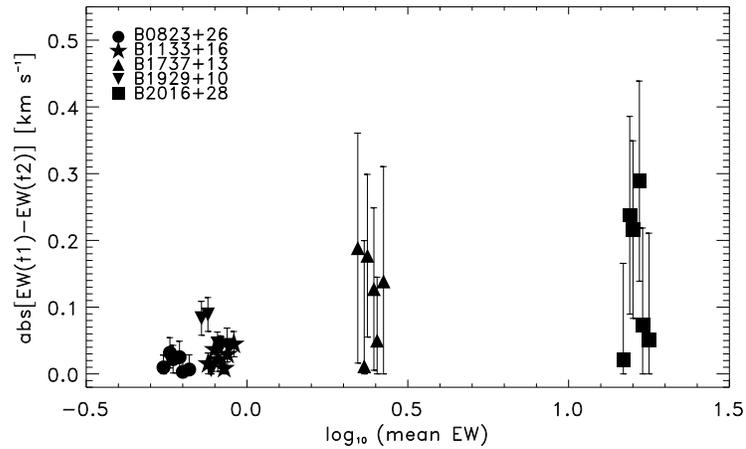}
\caption{\label{f:like_fig4}The absolute value of $\Delta {\rm EW}$ as
a function of the mean {\rm EW} for each pulsar,
$\overline{\rm EW}$. Plotted error
bars represent $\pm$ 1-$\sigma$ uncertainties.}
\end{figure*}

We have measured the equivalent width
\begin{equation}
{\rm EW} = \int \tau dv = - \int \ln(I/I_{0}) dv,
\end{equation}
in units of \kms, for each observing epoch.
The optical depth profiles were integrated
over the `on-line' velocity regions.
To avoid low signal/noise parts of absorption spectra, as well as line wings which may be 
affected by small baseline imperfections, we have defined the `on-line' regions as roughly 
encompassing velocities with $\tau(v)>$2-$\sigma$.
For each pulsar a single region was used for all observing epochs:
$[-10, 15]$ \kms~for B0540+23, $[2, 7]$ \kms~for B0823+26,
$[-7, 0]$ \kms~for B1133+16, $[2, 6]$ \kms~for B1737+13, $[-3, 12]$ \kms~for B1929+10 and
$[-2, 21]$ \kms~for B2016+28. In addition,
by averaging optical depth profiles for all epochs we have estimated the mean equivalent 
width for each pulsar ($\overline{\rm EW}$).

To compare {\rm EW} from one epoch to the next we have calculated
$\Delta {\rm EW}(\Delta t) = [{\rm EW}(t_{1}) - {\rm EW}(t_{2})]$.
These values, as a function of time baselines $\Delta t$ (bottom axis) or
traversed spatial scales $l$ (shown on the
top axis), are shown for all pulsars in Figure~\ref{f:comparison_EW}.
B0540+23 was excluded from this analysis as its optical
depth profile is uncertain around 0 \kms.
To convert from time intervals $\Delta t$ to spatial scales $l$, we use $l=\Delta t \times L_{\perp}$, 
where $L_{\perp}$ is provided in Table 1.
Our spatial scales given below should be multiplied
by an (unknown) fractional distance factor $f = D_{\rm HI}/D_{\rm PSR}$ to
account for the fact that absorbing HI is located between us and the pulsar.
Note that some authors have arbitrarily fixed $f$; e.g.
for \cite{Minter05} $f=0.5$. (When we assemble various workers' results in this
paper, we redefine all such values onto our scale.)

As shown in Figure~\ref{f:comparison_EW},
in the case of  B0823+26, B1133+16, B1737+13 and B2016+28 all variations in 
$\Delta {\rm EW}$ are essentially within  2-$\sigma$ and are most likely not significant.
However, in the case of B1929+10, baselines of 0.33 and 1.24 yrs show changes 
in $\Delta {\rm EW}$ at a $3.3$-$\sigma$ and $3.5$-$\sigma$ level, respectively.
In addition, the 0.16 yr baseline shows a variation at a $2.8$-$\sigma$ level, while 
the 0.75 yr baseline shows a 2.5-$\sigma$ change.
To quantify the significance of $\Delta {\rm EW}$ variability we test the null hypothesis 
of $\Delta {\rm EW}$ being constant with time. We fit a constant $\Delta {\rm EW}$ model 
to measurements in Figure~\ref{f:comparison_EW} and find $\chi^2/{\rm dof}=$0.4
(B0823+26), 0.8 (B1133+16), 1.3 (B1737+13),
6.0 (B1929+10), and 1.2 (B2016+28).
We conclude that $\Delta {\rm EW}$ changes for B1929+10 can not be fit with a 
constant model, while for all other pulsars no
significant variability of $\Delta {\rm EW}$ is observed.
This $\Delta {\rm EW}$ variability, as further analyzed in Section 7.1, can be interpreted as being due to TSAS.

In Figure~\ref{f:like_fig4} we plot $\Delta {\rm EW}$, separately for individual 
pulsars, but as a function of $\overline{\rm
EW}$.  This plot can be directly compared with Figure 4 from Frail et
al. (1994). Our results
for B0823+26, B1133+16 and B1929+10 agree roughly with Frail et al.'s, $\mid \Delta {\rm EW} \mid <0.1$.
However $\Delta {\rm EW}$ for B1737+13 and B2016+28 found in Frail et
al. (1994) were significantly larger than what we find. A particularly
striking difference is the case of B2016+10,
where Frail et al. (1994) found $\Delta
{\rm EW} \sim1$--5, while we find only $\Delta {\rm EW} \sim 0.1$--0.3.
We believe that at least some of the variations Frail et al. saw were due to
small calibration errors.
Looking at our whole sample,
the fractional variation $\Delta {\rm EW}/\overline{\rm EW}$ spans a range from 0\% to 11\%, with a median of 4\%. 
This is significantly smaller from a median value of 13\% found in Frail et al. (1994).

As a conclusion, we find significant variations in $\Delta {\rm EW}$ only for B1929+10, with two largest 
changes being at a fractional level of 11\%.
The frequency of variations and the level of variations (as well as upper limits)
are significantly smaller than what
was found previously for the same sources  by Frail et al. (1994). Our results do not rule out the 
finding by \cite{Frail94} that
$\mid \Delta {\rm EW} \mid$ increases with EW.
However, if this relationship exists, it is at a much smaller level than suggested by Frail et al.

\section{Variations from difference spectra}
\label{s:results}

Considering that TSAS may have temperature as low as 15 K (Heiles 1997), the change in absorption
profiles may occur over $\la 1$ \kms~and appear in only one or two
velocity channels in our spectra. Such narrow changes may get washed out
when calculating the equivalent width.
We therefore in this section
compare all pairs of absorption profiles for a given pulsar, and investigate absorption 
spectrum differences, $\Delta (I/I_{0})\sim \Delta \tau$.
With four observing epochs we have six difference spectra (Figures~\ref{f:0823_6plot} 
to \ref{f:2016_6plot}). Overlaid atop each difference spectrum is a  $\pm$2-$\sigma$ 
significance envelope calculated from noise envelopes discussed in
Section~\ref{s:noise}.

In summary, we see changes in difference spectra at a 3-$\sigma$ level only in the case of B1929+10
and only for the two baselines where we also find changes in $\Delta {\rm EW}$ at a $>3$-$\sigma$ level.
We note that B1133+16 has one feature above 3-$\sigma$ (s4-s3 baseline) at a velocity of $-30$ \kms, 
however this likely comes from RFI as is seen in {\it emission} in the absorption spectrum from session 
3. Similarly, B1737+13 has a significant variation at a velocity of $\sim35$ \kms on 3 baselines, all
involving the session 1 absorption spectrum; this corresponds to an (unphysical) 
{\it emission} feature and is most likely due to RFI.

For three of our pulsars (B0823+23, B1133+16 and B1929+10) we have excellent sensitivity 
for TSAS experiments, $\Delta \tau\sim0.01$-0.02 (2-$\sigma$ in the on-line portion of the spectrum).
Only one recent experiment (Minter at al. 2005) reached similar sensitivity but
primarily over smaller spatial scales.
For B1737+13 and B2016+28 we have reached 2-$\sigma$ sensitivity of 0.1-0.2. Although this 
is slightly higher, several pulsar and
VLBA TSAS detections have been claimed at this level.
We provide below brief comparison between our observations for individual 
pulsars and previous TSAS experiments.

\subsection{Difference spectra of individual sources}

\subsubsection{PSR B0540+23}

The corresponding tranverse distances traveled by B0540+23 during the
course of our observations are 13, 26, 60, 73, 86 and 100 AUs.
Frail et al. (1994) concluded that there
were clear signs of significant optical depth variations in some absorption
components for this pulsar. In fact both our and Frail et al.'s
spectra show ($I/I_{0}$) slightly less than zero near 0 \kms~as well as scintillation ripples. 
As already discussed in Section 3.2, we do not use data for this pulsar for our
further analyses and do not show difference spectra.

\begin{figure*}
\plotone{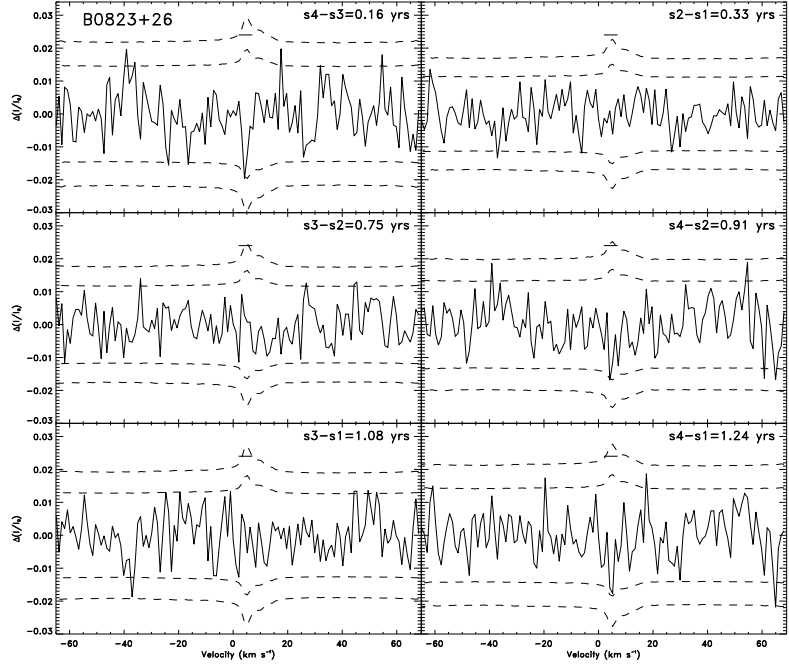}
\caption{\label{f:0823_6plot} Difference spectra for B0823+26
derived from four different observing epochs. Dashed lines show the expected $\pm$2- and 3-$\sigma$ noise
envelopes, whose shape reflects the HI emission in the beam.
The vertical scale is in the original units of the absorption spectra,
$\Delta(I/I_{0})$, which, for small $\tau$, is also equal to
$\Delta(\tau)$. The `on-line' region is shown with a horizontal bar.
}
\end{figure*}

\subsubsection{PSR B0823+26 (Figure~\ref{f:0823_6plot})}

With   $L_{\perp}=39$ AU yr$^{-1}$, the time intervals
covered by our experiment translate to spatial scales of 6, 13, 29, 35, 42 and 48 AUs.
There are no data points with $\Delta \tau$ above 3-$\sigma$ in difference spectra for all six time baselines.
We conclude that {\em{no}} significant change in absorption profiles is found down to a $\Delta\tau$ level of 0.03
(a 2-$\sigma$ level on \hi\ line). Frail et al. (1994) detected only marginal changes over 0.6 yr, but reported
significant variations of $\Delta \tau \sim0.07$ in 1.1 yr. We do not confirm this reported 
variation on any of our temporal
baselines although our 2-$\sigma$ sensitivity level  on all
baselines is at least two times smaller than the level of variations found by Frail et al. (1994). 
We discussed this further in \cite{Stanimirovic03b} and concluded that calibration and noise 
issues affected the Frail et al. analyses.

\begin{figure*}
\plotone{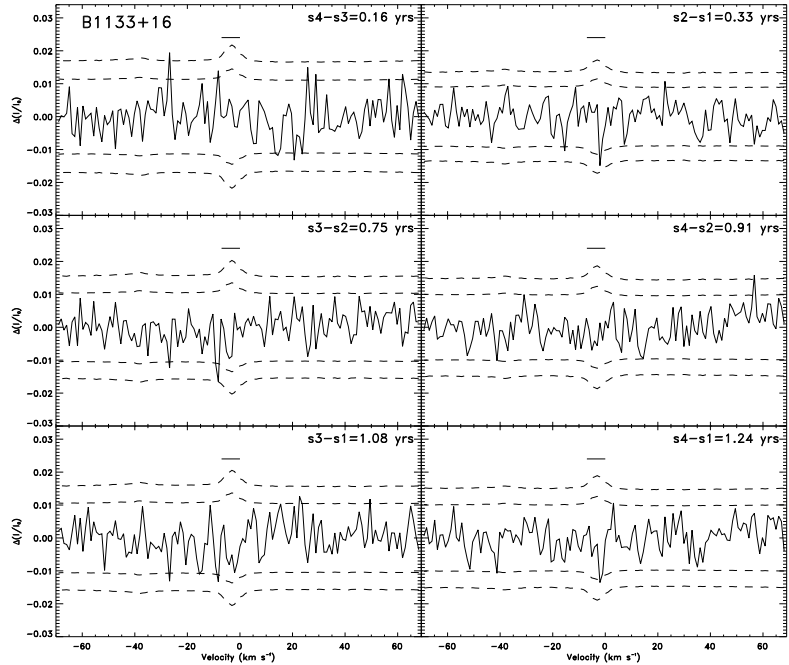}
\caption{\label{f:1133_6plot}Difference spectra for B1133+16
derived from four different observing epochs. See Fig.~\ref{f:0823_6plot} for further details.}
\end{figure*}

\subsubsection{PSR B1133+16 (Figure~\ref{f:1133_6plot})}

As this pulsar has $L_{\perp}=135$ AU yr$^{-1}$ a wide range of spatial scales is
probed by our observations, from 20 AU to 170 AU.
No variability at $\ga3$-$\sigma$ was found and
we conclude that {\em{no}} significant variations were detected
in the direction of this pulsar
down to a $\Delta\tau$ level of about 0.02.
Frail et al. (1994) reported $\Delta \tau \sim 0.04$ on a
1.1 yr baseline.
On all of our difference spectra (time baselines of 0.33, 0.75 and 1.08 yr)
our 2-$\sigma$ sensitivity is half the level of variations
found in Frail at al. (1994), $\Delta \tau \sim0.02$ at most.
We note that no variability was detected at this level in any of later variability experiments.

\begin{figure*}
\plotone{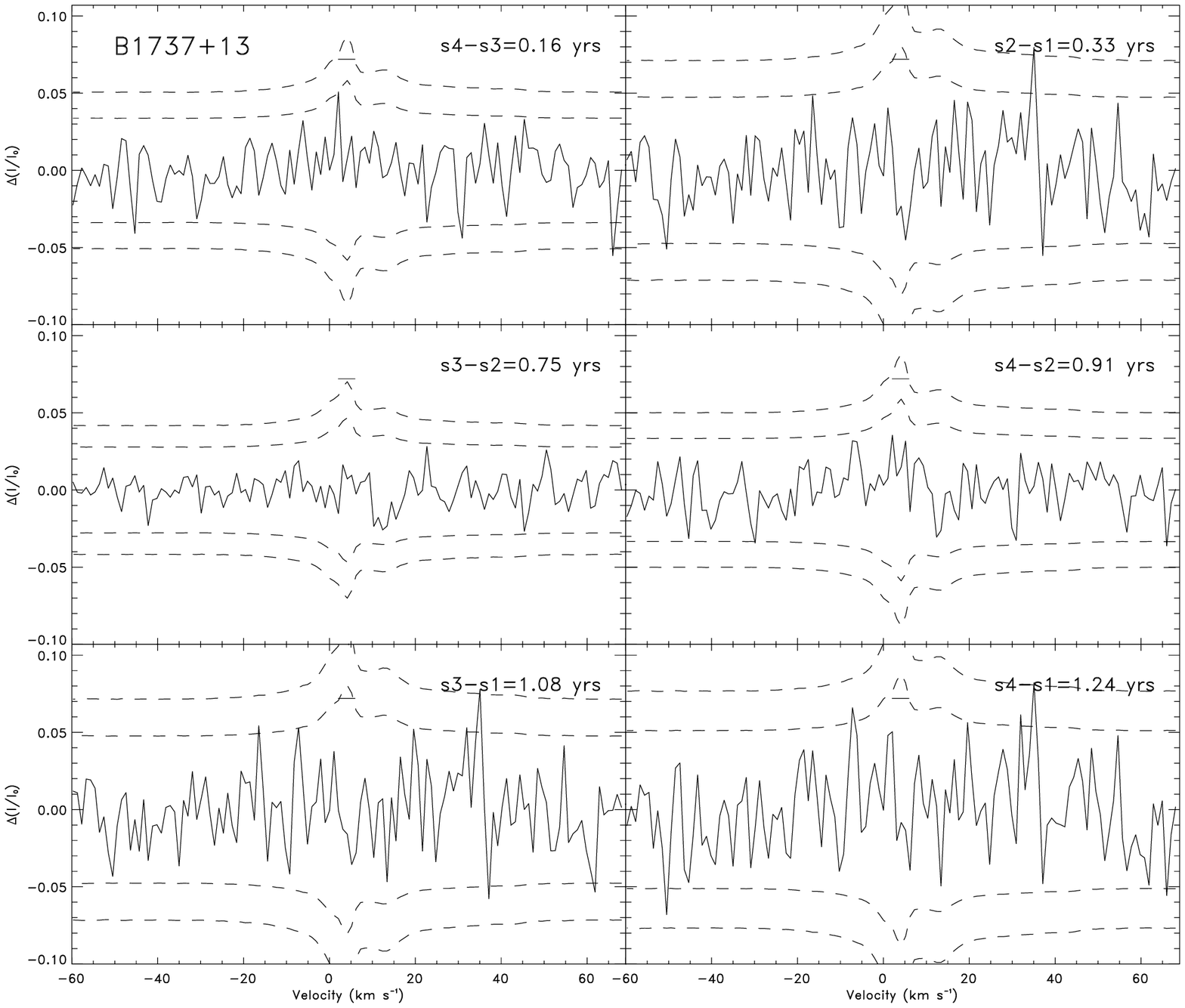}
\caption{\label{f:1737_6plot}Difference spectra for B1737+13
from four different epochs. See Fig.~\ref{f:0823_6plot} for further details.}
\end{figure*}

\subsubsection{PSR B1737+13 (Figure~\ref{f:1737_6plot})}

With $L_{\perp}=142$ AU yr$^{-1}$ the range of probed spatial scales
is from 20 AU to 175 AU.
{\it No} significant variations are found.
Frail et al. (1994) reported $\Delta\tau=0.5$ at 4 \kms (which corresponds to
$\Delta(I/I_{0}) \sim 0.1$ in this relatively deep line) over 0.65
yr. Our 2-$\sigma$ sensitivity is $\Delta(I/I_{0}) \sim0.2$ for half of baselines, and 0.3 for the rest.
Several TSAS features were detected at a similar level in recent experiments, both pulsar
and interferometric.

\begin{figure*}
\plotone{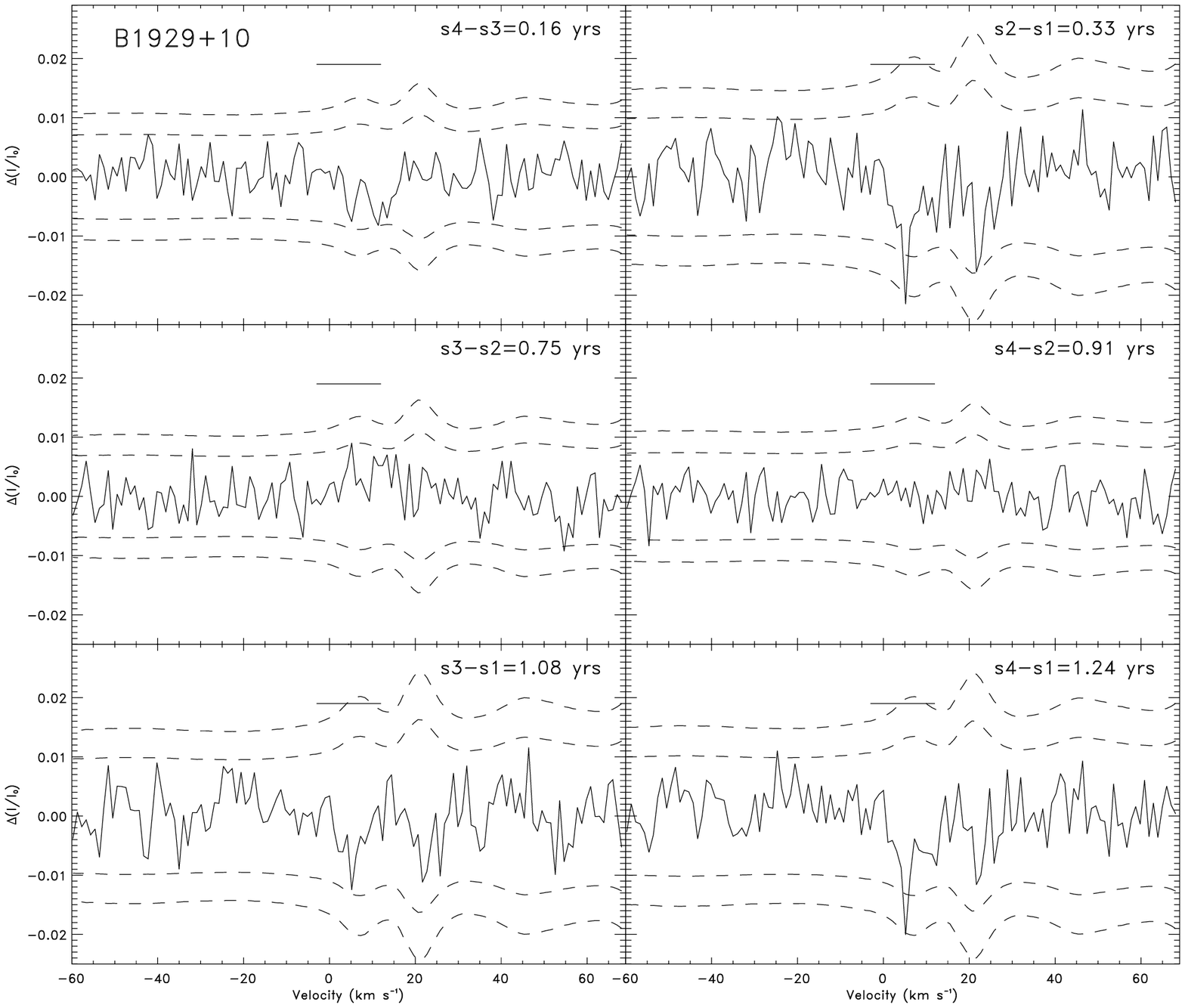}
\caption{\label{f:1929_6plot} Difference spectra for B1929+10 derived from four different 
observing epochs. Dashed lines show the expected $\pm$2- and 3-$\sigma$ noise envelopes, 
whose shape reflects the HI emission in the beam plus uncertainties associated with ghost 
excision (See \S{\ref{s:baseline}}).}
\end{figure*}

\subsubsection{PSR B1929+10 (Figure~\ref{f:1929_6plot})}

\begin{figure*}
\epsscale{0.8}
\plotone{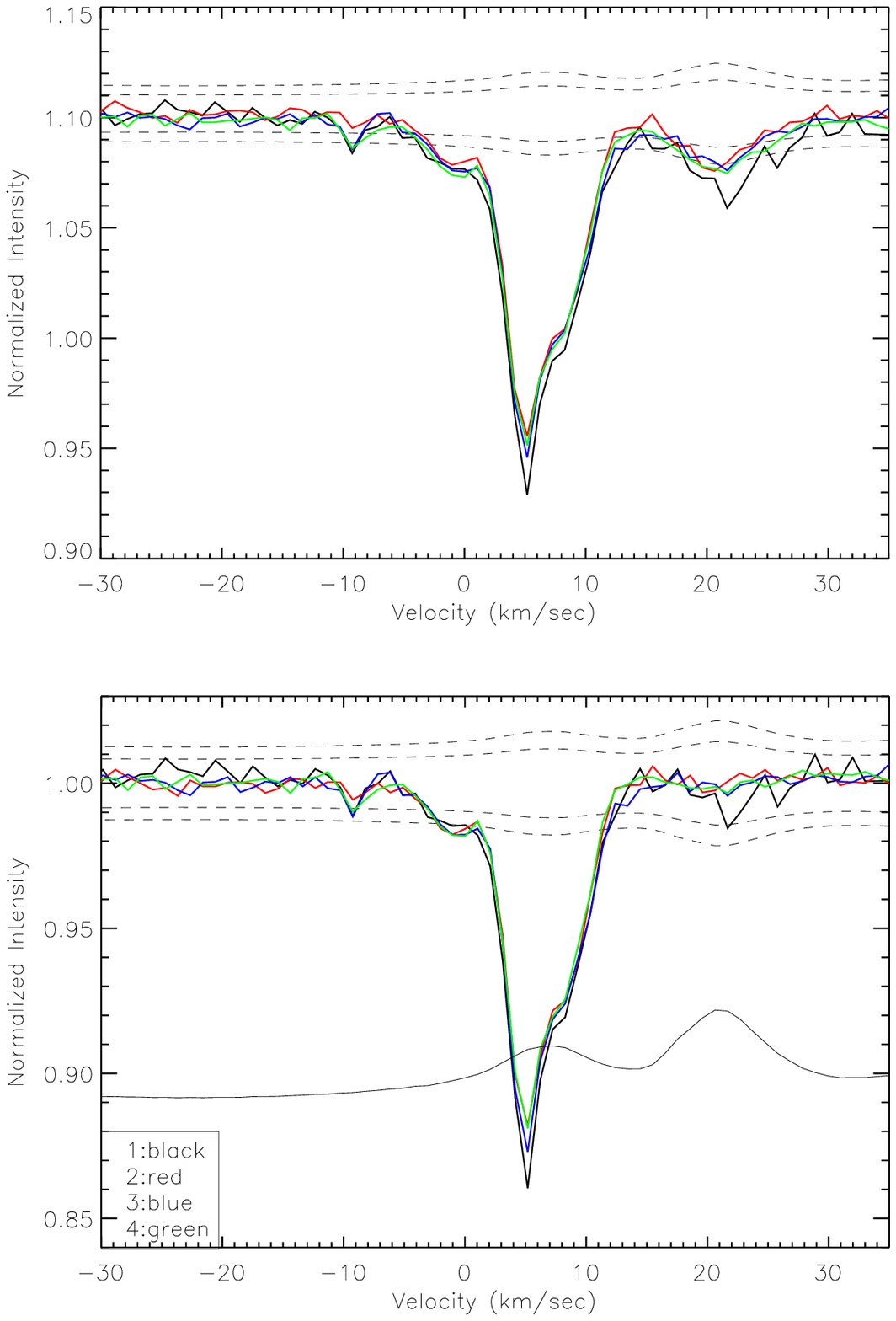}
\caption{\label{f:1929_4plot} HI absorption spectra for B1929+10 from all four epochs
before (top) and after (bottom) ghost removal.
Dashed lines show the expected $\pm$2- and 3-$\sigma$ noise envelopes for the 1st epoch spectrum
The HI emission spectrum, which was exorcized from the displayed absorption spectra
(see text), is shown as a faint black line at normalized intensity $\sim0.92$.
}
\end{figure*}

With $L_{\perp}=37$ AU yr$^{-1}$, the range of probed spatial scales is
very narrow, from 6 to 45 AU.  The absorption profile for this pulsar consists of at
least three different velocity components (at $-1.5$, 5 and 8 \kms).
The component at 5 \kms, which is the one with the deepest HI
absorption, is the only one exhibiting significant changes.
Difference plots show line variations at 3.2 and 3.0-$\sigma$ levels ($\Delta\tau=$0.025)
for time baselines of 0.33 and 1.24 years.
We consider these detections to be significant, especially as
the same time baselines showed significant variations in EW.
This level of variability is the lowest ever detected in TSAS experiments.
Both detected difference features are sharp and appear in the same single velocity channel at
5 \kms~over different epochs. We further measure and analyze properties of the
intervening absorbing
clouds traced by these detected features in
Section~\ref{s:sig_features}. Baselines of 0.16, 0.75, 0.91, and 1.08 years have an
upper-limit in $\Delta \tau$ of 0.012-0.018 at this velocity. Frail et al. noticed variations of up to  
$\Delta\tau\sim 0.03$ for this pulsar,
with the largest fluctuations being at a velocity of $\sim8$ \kms, where we observe no significant variations.

As this is the only pulsar in our sample to exhibit significant 
absorption variations, we have scrutinized the data 
acquisition and analysis procedure with extra care. For example, a scaling error of approximately
$10 \%$ in pulsar strength in a single epoch could result in an 
erroneous rescaling of optical depths by a similar 
amount, which would lead to a spurious detection of variation when there is actually none. For example, in 
Stanimirovic et al (2003), we pointed out an apparent case of 
this type of problem in Frail et al's spectra of PSR 
B2016+28. However, it appears that the more
advanced CBR spectrometer is indeed able to measure spectra 
more accurately, as we anticipated. For example, we 
see no evidence of variations in our experiment with 
PSR B2016+28 nor with B1133+16, the other particularly bright 
pulsars in our sample.  (Calibration errors would be expected to 
be most serious in the brighter pulsars, which can 
easily increase the system temperature by factors of more than two.)

It is notable that B1929+10 is the only pulsar in our sample from which we removed a ghost. Then an obvious 
question is
whether our detected variability could be an artefact of the ghost fitting procedure.
We have scrutinized our reduction procedures and remain confident that the variability is real.
Spectra for B1929+10 were reduced in the same way as for all other sources. 
Even the ghost fitting procedure was 
done in exactly the same way (using exactly the same IDL routines) as the 
polynomial baseline fitting for other 
pulsars. The derived error envelopes
reflect additional uncertainties due to this fitting process.

In Figure~\ref{f:1929_4plot} we show all four absorption profiles for this pulsar
before (top) and after (bottom) ghost removal.
Only the deepest velocity component at 5 \kms~shows variability (both before
and after ghost removal), while components at $-1.5$ and 8 \kms~agree extremely well between epochs.
If there were some imperfections related either to a scaling problem or to 
ghost fitting errors, they would affect {\it{all}} 
features. We have also examined the estimated coefficients used in equation
(2). In the case of the 1st session spectrum, for example, $A=-4.3 \pm 0.4$.
A smaller value for $A$ would result in a larger correction and would bring 
the 1st session absorption spectrum closer 
to spectra from other sessions. However,
even by taking values within $\pm 2-3\sigma$ of the best-fit estimate,
the observed variations in the difference spectra would still stay significant
(above 3-$\sigma$). The reason for this is that the HI emission 
at 5 \kms~is not exceptionally high, resulting in a small 
ghost correction.
After all these tests, we remain confident that the detected variability
in the case of B1929+10 is real.

\begin{figure*}
\epsscale{1.2}
\plotone{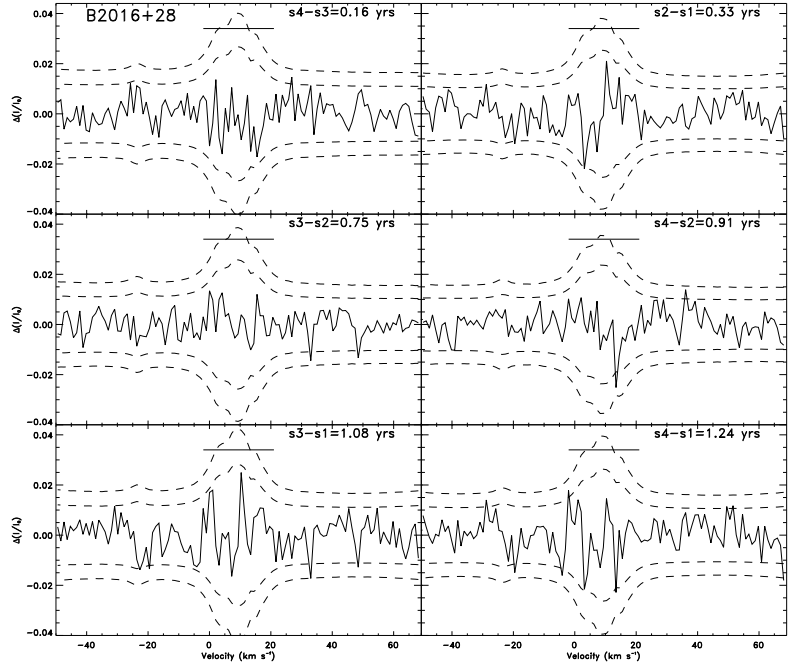}
\caption{\label{f:2016_6plot}Difference spectra for B2016+28
derived from four different observing epochs. See Fig.~\ref{f:0823_6plot} for further details.}
\end{figure*}

\subsubsection{PSR B2016+28 (Figure~\ref{f:2016_6plot})}

In Figure~\ref{f:2016_6plot} we  compare  all four absorption
profiles for this pulsar. This is the slowest-moving pulsar in our sample, so the range of spatial scales
covered with our observations is only 1--10 AUs.
No significant variations were found. Our 2-$\sigma$ sensitivity is 
$\Delta(I/I_{0})\sim0.15$ for all baselines.
Frail et al. (1994)  found $\Delta(I/I_{0}) \sim 0.15$ (from their Fig. 3)
over periods of 0.6 and 1.7 yr. As first noted in 
Stanimirovic et al. (2003), we compared one of our absorption
profiles for this pulsar with data from Frail et al.
and concluded that slight calibration problems may have affected Frail
et al.'s data. Two recent pulsar experiments found variability at this level 
but at much larger scales ($>200$ AU; 
Johnston et al. 2003 and Weisberg et al. 2008).

\subsection{Additional Opacity Fluctuations in Unsmoothed Difference Spectra?}
\label{s:single}

\cite{Weisberg07} presented preliminary results of this experiment, and noted that the
unsmoothed HI absorption spectra exhibited several apparently
significant optical depth variations, mostly in the form of single-channel fluctuations.  The most
prominent ones were in the direction of B1929+10, but there were a few additional
variable features toward other pulsars.  For the
current study, however, where we have applied hanning smoothing
to ensure full independence between velocity channels, only the  detections  in the direction of B1929+10 
survive.  Could
the additional variations in the unsmoothed spectra be real?

The minimum expected temperature for the CNM in thermal equilibrium is 15K \citep{Spitzer78,Heiles07}. 
To achieve such a low temperature, dust grain heating of the HI
would need to be absent. At this temperature, the full-width, half-maximum (FWHM) of
an HI feature will be 0.8 km s$^{-1}$ even in the absence of turbulent broadening; for a less extreme 30K 
the corresponding width is 1.2   km s$^{-1}$. Using a large sample of CNM clouds, \cite{Heiles03b} deduced 
a relationship
between $T_k$, the temperature inferred from the measured linewidth, and the spin temperature
$T_s:  T_k  \sim   T_s(1+ M^{2.42})$, where $M$ is the sonic Mach number. For 196 out of 202 CNM 
components they found that $M >1$,
implying then that $T_k/T_s >1.2$. Therefore, even with the minimum expected turbulence,
the HI absorption line FWHM would be $\sim$ 0.9 km s$^{-1}$ at $T_s \sim 15$ K, or 1.3 km s$^{-1}$ at 
$T_s \sim 30$ K. So while we have rejected most of \cite{Weisberg07}'s variations because unsmoothed data 
can be misleading, there are also physical reasons to doubt reality of features with FWHM$<1$ \kms. 
 In contrast, our hanning smoothed resolution
appears to match the narrowest plausible TSAS features.

\section{Decomposition into Gaussian Components}
\label{s:gaussian_results}

In this section we derive the spin temperature and the CNM column density
for gas seen in absorption in the direction of our pulsars. We
employ the technique developed by \citet{Heiles03a}, which is based on fitting Gaussian components
to both HI absorption and emission spectra.
The purpose of this section is twofold:
(i)  our further data analysis requires
temperatures and column densities of associated CNM clouds; and
(ii) since the CNM absorption profiles are often  superpositions
of several components,
our original hope was to search for temporal changes in both the peak
optical depth and velocity centroids of all individual velocity components.
Unfortunately, since the Gaussian decomposition of absorption and emission
profiles is not unique (see discussion in \citet{Heiles03a}), our second goal was not achievable.

To analyze each pulsar, we first  selected the epoch with the lowest noise level. The HI absorption
and emission spectra were then decomposed into several
Gaussian profiles. We note that three of our six pulsars are at high Galactic latitudes and have simple
 line profiles, easily fitted with
one or two Gaussian functions. Optical depth profiles for low latitude pulsars B0540+23,
B1929+10, and B2016+28, (at $b=-3.32$, $-3.88$, and $-3.98$\degree, respectively),
are more complex but still with several obvious velocity components.

\subsection{Procedure for estimating the spin temperature $T_s$}

We summarize here very briefly the basic definitions and the procedure, see Heiles \& Troland (2003a) for 
further details. For each pulsar we have
the HI emission profile, $T_{\rm B}(v)$ (note that this is the
emission spectrum that would be observed in the absence of the pulsar)
and the HI optical depth spectrum, $I(v)/I_{0}=e^{-\tau(v)}$. Both spectra
in general have multiple velocity components.

We first fit $\tau(v)$  with a set of $N$ Gaussian functions
using a least-squares technique:
\begin{equation}
\tau (v) = \sum_{0}^{N-1} \tau_{0,n} e^{-[ (v-v_{0,n})/\delta v_{n} ]^2}
\label{eqn:tau}
\end{equation}
where $\tau_{0,n}$ is the peak optical depth,
$v_{0,n}$ is the central velocity, and $\delta v_{n}$ is the 1/e width
of component $n$. $N$ is the minimum number of components necessary to make the residuals of the 
fit smaller or comparable to the estimated
noise level of  $I(v)/I_{0}$.

While the optical depth spectrum predominantly  reflects the CNM, both the cold {\it and} warm 
neutral media (WNM) contribute to the HI emission spectrum.
Consequently we can write the amplitude of the emission spectrum,
$T_{\rm B}(v)$, as follows:
\begin{equation}
T_{\rm B}(v)= T_{B,CNM}(v) + T_{B,WNM}(v).
\label{eqn:Tb}
\end{equation}
We now write each of the two terms of Eq.  \ref{eqn:Tb} in detail,
using the  optical depth profile derived from Eq. \ref{eqn:tau}.

First, $T_{B,CNM}(v)$, the HI emission originating from $N$ CNM components,
can be written as:
\begin{equation}
T_{B,CNM}(v) = \sum_{0}^{N-1} T_{s,n}(1-e^{-\tau_{n}(v)})
e^{-\sum_{0}^{M-1} \tau_{m}(v)},
\end{equation}
where $T_{s,n}$ is the spin temperature of cloud $n$, and
the subscript $m$ represents one of the $M$ CNM clouds that lie
in front of cloud $n$.

Next, $T_{B,WNM}(v)$, the HI emission originating from the
WNM, is represented with a set of $K$ Gaussian functions. The complicating factor here is that a certain
 fraction $F$ of the WNM is
located in front of the CNM, while a fraction $(1-F)$
of the WNM is beyond the CNM with its emissions being absorbed by CNM clouds: \begin{equation}
T_{B,WNM}(v)= \sum_{0}^{K-1} [F_{k} + (1-F_{k})e^{-\tau(v)}] \times
T_{0,k}e^{-[(v-v_{0,k})/\delta v_{k}]^2 },
\end{equation}
where the subscript $k$ corresponds to each of the WNM components and a fraction
$F_{k}$ of the WNM cloud $k$ lies in front of all CNM components, while a
fraction $1-F_{k}$ is being absorbed by the CNM clouds.
For a given order of CNM clouds along the line of sight and a given set of $F_{k}$ values, the
$T_{\rm B}(v)$ profile is simultaneously fitted for Gaussian
parameters of the WNM components and the spin temperature of individual
CNM clouds.

For each pulsar, we vary the order of Gaussian functions along the
line-of-sight (for $N$ CNM components there are $N!$ possible orderings)
and perform the $T_{\rm B}(v)$ fit. We then choose the ordering of CNM components that gives the smallest
residuals in the least-squares fit.
Unfortunately, the
difference in the fit residuals is often not sufficiently statistically significant to
distinguish between different values of $F_{k}$. However, $F_{k}$ has
a large effect on the derived spin temperatures. Hence we follow the
Heiles \& Troland (2003a) suggestion and estimate the final spin temperatures by
assigning characteristic values of 0, 0.5, or 1  to each $F_{k}$ (among the extreme possible 
values of 0 and 1), and repeating this for all possible
combinations of WNM clouds. The final spin temperatures are then
derived as a weighted average over all trials.

\subsection{Spin temperature and HI column density}
\label{s:spin_temp}

\begin{figure*}
\plotone{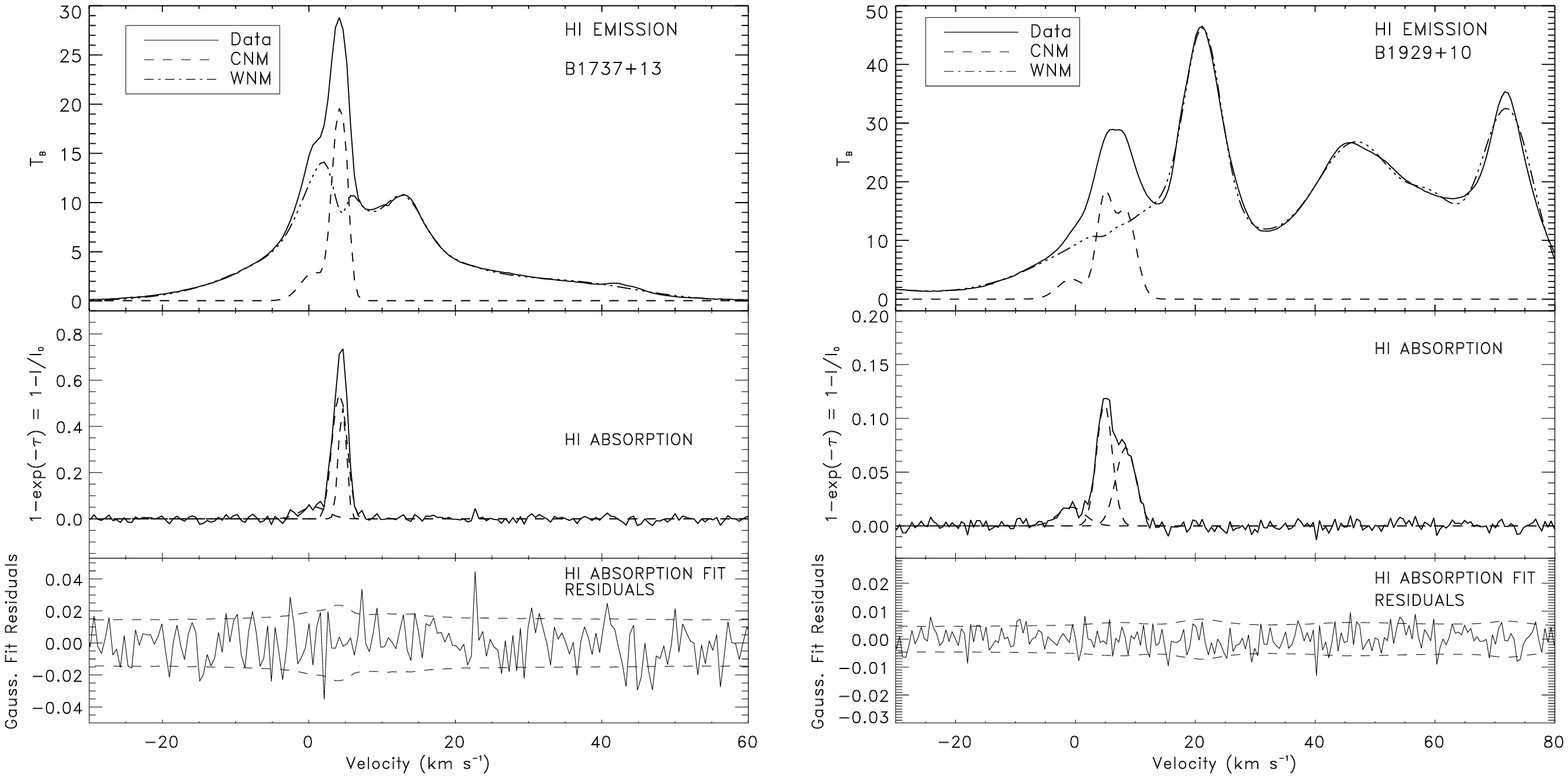}
\caption{\label{f:1737_wnm_cnm}
{\bf Left:} Fitted results for the opacity profile and \hi\
emission for B1737+13 observed at epoch 2001.86. {\bf Right:} Fitted results for
the opacity profile and \hi\
emission for B1929+10 observed at epoch 2000.95.
({\it Top})
The \hi\ emission profile is shown with the solid line. The total
\hi\ emission from WNM components is shown with the dot-dot-dashed line. The
dashed line shows the emission contribution from all
CNM components ($T_{B,CNM}(v)$).
({\it Middle}) The optical depth profile is shown with the solid
line. Fitted Gaussian components are shown with dashed lines.
({\it Bottom}) Residuals after fitting Gaussian function(s) to the
optical depth profile. The overlaid contours show $\pm$1-$\sigma$ levels.
}
\end{figure*}

Figure~\ref{f:1737_wnm_cnm} shows results from the
Gaussian decomposition and fitting of HI emission and absorption
spectra for two example pulsars, B1737+13 and B1929+10. This figure has three panels: the
{\it top panel}, displaying the \hi\ emission spectrum, $T_B(v)$, and its
decomposition into separate CNM [$T_{B,CNM}(v)$, shown as a
dotted line] and WNM [$T_{B,WNM}(v)$, shown as a dashed line] components; the
{\it middle panel} giving the optical depth profile and its
individual Gaussian-fitted components
(shown as dotted lines); and the {\it bottom panel} showing the
difference between the observed optical depth profile and its Gaussian
component fit (dashed lines on this plot show $\pm1\sigma$ noise levels). 
In Table~\ref{t:table_wnm_cnm} 
we list the fitted CNM properties.

With the exception of the B1929+10 line of sight, 
all derived spin temperatures listed in Table~\ref{t:table_wnm_cnm},
are in agreement with what is typically found for CNM clouds \citep{Heiles03b,Dickey03}.
For example,  \cite{Heiles03b} show histograms of WNM and CNM temperatures
from their HI emission/absorption survey of 79 continuum sources.
Most CNM clouds have a spin temperature in the range 20--70 K, while
CNM clouds with a temperature in the range 150--200 K appear to be rare.
However, spin temperatures we derived for CNM clouds in the direction of
B1929+10 are high, $\sim150$ and $\sim200$ K. This is the first indication
that this particular line of sight could be sampling atypical ISM conditions.

In Table~\ref{t:HI_column_density_for_PSRS} we provide the HI column density in WNM and CNM phases 
in the direction of each pulsar. The first three pulsars have the CNM/total column density of about 15-20\%, but
for B1929+10 this ratio is only 7\%, and for B2016+28 the ratio is about 40\%.
The very low CNM abundance in the direction of B1929+10 is unusual. For comparison, \cite{Heiles03b} found 
that all except one of their sources with detected CNM have a CNM fraction $>10$\%, with a majority of 
sightlines having $\sim30$\%.
However, the same study also noted a separate class of objects not exhibiting evidence for the CNM
along their lines of sight. They suggested that at least some of these regions
may be affected by supershells.
The CNM fraction in the direction of B1929+10 is similar to several directions with a CNM fraction of only
2--4\% observed by \cite{Stanimirovic05} and \cite{Stanimirovic07}.
Clearly, properties of the CNM in the direction of B1929+10 stand out as unusual.
As we will see in  \S\ref{s:1929_section}, a significant fraction
of this line of sight is located inside the Local Bubble.

\begin{table*}
\caption{CNM components from Gaussian decomposition. }
\centering
\label{t:table_wnm_cnm}
\begin{tabular}{lccccc}
\noalign{\smallskip} \hline \hline \noalign{\smallskip}
PSR      & $\tau_{0}$  & $v_{0}$ (LSR) & $\Delta v$ & $T_{\rm spin}$ &
N(HI)$_{\rm CNM}$ \\
    &   & (\kms) & (\kms) & (K) &
($10^{20}$ cm$^{-2}$) \\
\hline
B0823+26 & & & & \\
& $0.26\pm0.01$ & $4.91\pm0.02$ & $2.3\pm0.1$ &  $62\pm5$
& $0.72\pm0.02$ \\
\\
B1133+16 & & & & \\
& $0.16\pm0.02$ & $-2.10\pm0.03$ & $1.7\pm0.1$ &  $27\pm5$
& $0.15\pm0.01$ \\
& $0.16\pm0.01$ & $-3.5\pm0.1$ & $3.6\pm0.1$ &  $48\pm6$
& $0.53\pm0.03$ \\
\\
B1737+13 & & & & \\
& $0.05\pm0.01$ & $0.7\pm0.6$ & $4.2\pm0.8$ &  $59\pm6$
& $0.25\pm0.03$ \\
& $0.78\pm0.13$ & $4.19\pm0.06$ & $2.1\pm0.1$ &  $35\pm7$
& $1.1\pm0.1$ \\
& $0.66\pm0.13$ & $4.59\pm0.05$ & $1.2\pm0.1$ &  $22\pm6$
& $0.33\pm0.05$ \\
\\
B1929+10 & & & & \\
& $0.017\pm0.002$ & $-0.7\pm0.3$ & $5.3\pm0.8$ &  $199\pm25$
& $0.35\pm0.04$ \\
& $0.120\pm0.005$ & $4.8\pm0.1$ & $3.0\pm0.1$ &  $148\pm4$
& $1.00\pm0.04$ \\
& $0.076\pm0.003$ & $8.4\pm0.2$ & $3.9\pm0.3$ &  $206\pm7$
& $1.15\pm0.06$ \\
\\
B2016+28 & & & & \\
& $1.12\pm0.03$ & $3.70\pm0.03$ & $2.36\pm0.06$ &  $50\pm30$
& $2.8\pm0.3$ \\
& $0.67\pm0.04$ & $5.97\pm0.04$ & $1.4\pm0.1$ &  $50\pm30$
& $0.9\pm0.1$ \\
& $0.81\pm0.05$ & $9.16\pm0.06$ & $10.2\pm0.2$ &  $80\pm50$
& $13.3\pm1.5$ \\
& $0.57\pm0.05$ & $9.70\pm0.04$ & $2.7\pm0.2$ &  $50\pm30$
& $1.7\pm0.2$ \\
& $1.30\pm0.03$ & $13.3\pm0.01$ & $1.54\pm0.03$ &  $20\pm10$
& $0.7\pm0.1$ \\
\tableline
\end{tabular}
\end{table*}
\begin{table*} \caption{HI Column densities in CNM and WNM toward pulsars } \centering
\label{t:HI_column_density_for_PSRS}
\begin{tabular}{ccccc}
\noalign{\smallskip} \hline \hline \noalign{\smallskip}
PSR   & $N({\rm HI})_{\rm WNM}$ &$N({\rm HI})_{\rm CNM}$ & $N({\rm
HI})_{\rm TOT}$ & $N({\rm HI})_{\rm CNM}/N({\rm HI})_{\rm TOT}$\\
    & (10$^{20}$ cm$^{-2}$) & (10$^{20}$ cm$^{-2}$)& (10$^{20}$ cm$^{-2}$)& \\
\hline
B0823+26 & 4.3 & 0.7 & 5.0 & 0.14 \\
B1133+16 & 3.5& 0.7 &4.2 & 0.16\\
B1737+13 & 6.3 & 1.7& 8.0& 0.21\\
B1929+10 & 34.7& 2.5& 37.2&0.07\\
B2016+28 & 28.9& 18.7& 47.7&0.39\\
\noalign{\smallskip} \hline \noalign{\smallskip}
\end{tabular}
\end{table*}

\section{A Study of the Lines of Sight}
\label{s:1929_section}

We have examined the lines of sight of all pulsars in our sample. While B1737+13 and B2016+28 are at 
a significant distance from the Sun
and their lines of sight probably average over many interstellar clouds,
B0823+26, B1133+16 and B1929+10 are close with significant portions of their lines of sight probing the 
very local ISM.
We detail here some interesting properties of the latter three pulsars' lines of sight.

PSR B1929+10 is the closest pulsar in our sample, $D =  361^{+8}_{-10}$ pc \citep{Chatterjee04},
and this relatively close distance raises the question of whether the observed fluctuations of HI 
optical depth are somehow related to the pulsar's proximity to the solar system.  For example, we
are embedded near the middle of a low-density, high-temperature region called the
Local Cavity which extends 1-200 pc before being bounded by a high density wall in most directions, 
so a significant portion
of the line of sight lies inside it. In what follows, the ``Local Bubble (LB)'' consists of the Local Cavity plus the wall.
\citet{Lallement03} mapped the dense neutral gas wall bounding the Local Cavity via interstellar absorption
measurements of stars of known distance.  The updated map of the neutral gas surrounding the Local Cavity 
by \citet{Welsh10a} confirms that the line of  sight from B1929+10 enters the wall of the LB at a distance of 
$\sim120$ pc and then pierces the wall
for $\sim50-60$ pc. A lower-resolution interstellar reddening map from \citet{Toscano99} also indicates that 
the LB boundary is distended significantly toward the pulsar.
Furthermore, the B1920+10 line of sight  appears to be in or near a long extended neutral finger of the LB 
(possibly an interstellar tunnel gouged by a supernova)
and may therefore lie in the finger or along its wall for some distance. The dense finger is most likely 
associated with the Ophiuchus
molecular cloud and \cite{deGeus90} found that the CO distribution
in this region is filamentary and has radial velocities in the range 4.2-5.3 \kms (note that our varying HI 
component is centered at $4.8\pm0.1$ \kms).

In addition to the above, we found that
two stars  within 3 degrees from B1929+10 have been observed in NaI:
HD178125 is at a distance of 173 pc, while HD180555 is at a distance of only 106 pc. 
Based on the data presented in \cite{Genova97}, these stars show NaI absorption at an LSR velocity of
$+5.9$ and $+5.6$ \kms, respectively (velocity uncertainty for these
measurements is 1-1.5 \kms). As our varying pulsar absorption component is centered at 
$+4.8$ \kms~(with a FWHM of 3 \kms), it is most likely tracing the same interstellar cloud 
seen in the direction of the two HD stars. This would imply that
the absorbing cloud is at a distance of $<106$ pc, as it is seen against HD180555, and 
therefore embedded in the Local Cavity or the LB wall.

The Local Cavity itself is a region of low scattering; while most investigations have concluded
that  there is  enhanced  scattering at its boundary, implying enhanced turbulence
there \citep{Phillips92,Bhat98,Gupta99,Chatterjee01}.
Based on modeling of many NaI absorption features, \cite{Genova97} found that
the $+5.9$ and $+5.6$ \kms~absorption components in the direction of HD178125 and HD180555 
belong to a plane-parallel flow immersed in the Local Cavity, the so-called ``flow C''.
They suggested that flow C could represent gas condensing out of the LB wall. Generally, many 
diffuse interstellar clouds (so-called
``Local Fluff'')  have been found inside the Local Cavity itself  and have
been interpreted as signatures of the development of  Rayleigh-Taylor instabilities  in the wall, 
whereby gas is condensing
out of the leading edge of the LB wall and flowing toward the Sun \citep{Muller06,Breitschwerdt06}.

To examine lines of sight toward B0823+26 and B1133+16,
we turn to Figure 7 of \citet{Lallement03} which shows neutral gas in
vertical planes at fixed galactic longitudes.
In the case of B0823+26 ($b\sim32\degr$), about one-half of the line of sight appears 
to be inside the Local Cavity and  traverses the LB wall over only a short distance. In the case of 
B1133+16 ($b\sim69\degr$), the line of sight appears to exit the LB through an opening or a 
chimney of low density material rather than a bounding wall or
any other substantially dense neutral material.

In summary, since  almost one-half of the line of sight toward B1929+10 is associated with the LB
and the absorbing gas is likely to be within 106 pc,
we hypothesize that our detected TSAS clouds along this line of sight are formed or influenced
by the LB.
Based on the stellar NaI absorption lines at similar velocities to those we find
in HI toward B1929+10, we conclude that our TSAS fluctuations are likely to be associated with the LB wall. 
The high spin temperature of the absorbing clouds, as shown in
the previous section, may result from the thermally unstable
gas due to wall fragmentation (see also Section 9.4).
It is then not surprising that we did not detect
TSAS in the direction of B0823+26 and B1133+16, despite equally good sensitivity, because these
lines of sight  do not traverse much or any of the wall.

\section{TSAS properties inferred from observations}
\label{s:sig_features}

In Sections~\ref{s:EW} and \ref{s:results} we have identified several significant variations in absorption 
profiles that could be interpreted as TSAS. All of these features are in the direction of B1929+10.
We summarize their observed properties in Table~\ref{t:tsas_sig_features}.
We measure TSAS properties from EW variations, as this is more
robust than using difference spectra which have a lower S/N.
We estimate TSAS properties from EW variations with $\ga2.5$-$\sigma$.

Table~\ref{t:tsas_sig_features} lists the following quantities:
the time baseline over which the change in EW is observed, the corresponding transverse 
spatial scale, the absolute value of $\Delta {\rm EW}$, the significance of the fluctuation,
the HI column density of the TSAS ($N({\rm HI})_{\rm TSAS}$),
the inferred HI volume density of TSAS (assuming
spherical HI blobs whose radius is equal $l$), and the inferred thermal pressure. Some comments 
are necessary for interpreting these tables.
(i) Our observations do not constrain the exact {\em{size}} of 
TSAS, but rather its characteristic {\em{scale}}.
In  measuring the tranverse distance ($l$) traveled by the pulsar
over a time span exhibiting a variation in HI absorption spectra, we are sampling a spatial scale 
over which we have sampled the optical
depth function. Therefore, even in the case that significant fluctuations are due to discrete blobs 
of gas, $l$ does not necessarily correspond to
a physical size of these blobs. This point has  been emphasized by Deshpande (2000).

(ii) The temperature of TSAS, needed to estimate $N({\rm HI})_{\rm TSAS}$, is also not 
constrained directly by observations.
We have estimated in Section 6 the temperature of the parent CNM clouds with which TSAS
is associated.
In the scenario where TSAS results from discrete, elongated objects, such as in
\cite{Heiles97} and \cite{Heiles07b}, CNM clouds are constituted from TSAS and an inter-TSAS medium.
The derived spin temperature in this case corresponds to the inter-TSAS medium, while
TSAS could have a much colder temperature of $\sim15-30$ K (Heiles 1997).

There are several interesting results:
\begin{enumerate}
\item Significant fluctuations (Table~\ref{t:tsas_sig_features}) are found on a variety of spatial scales 
$l= \frac{D_{\rm HI}}{350~{\rm pc}} \times (6, 12, 28, 46)$ AU (with $D_{\rm HI}$ being a TSAS distance), 
suggesting that the CNM in the direction of B1929+10 has structure on spatial scales of 5--50 AUs.
Our characteristic scales for TSAS are
similar to results from VLBA experiments:
Brogan et al. (2005) found a typical size of 25 AU and Lazio et al (2009)
found a typical size of 10 AU.

\item If we assume that TSAS has the same spin temperature as its parent CNM cloud ($\sim170$ K
in the case of B1929+10),
we estimate a HI column density of $N({\rm HI})_{\rm TSAS}=(1-3)\times 10^{19}$ cm$^{-2}$.
This is slightly lower than the typical CNM column density of $5\times10^{19}$ cm$^{-2}$ found 
for the CNM in the Heiles \& Troland (2003b) survey.

\item All TSAS features contribute 10-30 percent to the HI column
density of their `parent' CNM clouds.

\item If we interpret TSAS as being in the form of spherically symmetric blobs, its observationally inferred HI volume
density and thermal pressure are $>10^{4}$ cm$^{-3}$ and  $> {\rm a~few} \times 10^{6}$ cm$^{-3}$ K, respectively.
These values are more than two orders of magnitude larger than what is typically
expected for the ISM, in agreement with previous TSAS measurements
(see Section 2).

\item
In terms of HI mass, TSAS contains only
$\sim10^{-3}$ Earth masses. This is
many orders of magnitude below what is required
for gravitationally confined clouds.

\item
In the shaped TSAS model (Heiles et al. 1997) 
TSAS could be much colder with a temperature up to a factor of 10
lower than what we have used.
This would reduce the derived $N({\rm HI})_{\rm TSAS}$ and $n({\rm HI})$ by a 
factor of 10, and $P/k$ by a factor of 100. If an elongation factor
$G\sim5-10$ is further introduced, the derived thermal pressure can be brought down
to $P/k\sim2\times10^{4}$ cm$^{-3}$ K which is close to the traditional hydrostatic 
pressure (with some turbulent contribution).
However, this could work only for the case of {\it both} low TSAS temperature and significant
line-of-sight elongation.

\end{enumerate}

In Table~\ref{t:tsas_upper} we provide upper limits on 
TSAS properties based on non-detections in the directions of four other pulsars.

\section{Discussion}
\label{s:discussion}

\subsection{The Heiles (1997) discrete cloud model}
\label{s:heiles}

In the Heiles (1997) scenario, a typical CNM cloud consists of an inter-TSAS medium and many 
discrete TSAS features in the form of cylinders or disks, which are homogeneously and
isotropically distributed within the cloud. An arbitrary line of sight
samples many cylinders or disks; however only cylinders seen end-on
or disks seen edge-on have sufficient column densities to produce absorption features recognized as
TSAS. Meanwhile, non-end-on cylinders or non-edge-on disks have very little
column density and blend to form a smooth background
which contributes very little to the measured HI absorption profile.
The model assumed the following basic parameters for TSAS:
a cloud size perpendicular to the line of sight  $L_{\perp}= 30$ AU, $T_{\rm s}=15$ K,
$N({\rm HI})_{\rm TSAS}=0.33\times10^{19}$ cm$^{-2}$, and
$\aleph=3$ for the number of end-on cylinders or edge-on disks along every line of sight. Two
values were considered for $r$, the ratio of the CNM to TSAS column density:
$r=N({\rm HI})_{\rm CNM}/N({\rm HI})_{\rm TSAS} = 30$ and 100. The model predicts
the elongation factor $G$ and the volume filling factor $\phi$
of TSAS required to ameliorate the over-pressure problem, under the assumption
that the asymmetric TSAS features are uniformly and isotropically distributed throughout CNM clouds.
If TSAS is in the form of edge-on disks then $G\sim10$ and they fill about
3.3\% of the CNM volume; whereas if TSAS is in the form of end-on cylinders then $G\sim3.8$ 
and they fill about 3.6\% of the CNM volume.
However, the volume filling factor of TSAS is directly proportional to
the number of end-on cylinders or edge-on disks $\aleph$ along any line of sight.

If we assume $T_s=15$ K for our TSAS detections, a geometric factor
$G=5-10$ is required to explain the observed TSAS volume density and pressure  within the 
context of the Heiles model. This is in the range proposed by Heiles (1997). In addition,
the TSAS HI column density as well as the $r$ ratio are similar to values explored in the model. 
It is therefore reasonable to compare our estimated volume filling factor of TSAS
with the model predicted values.

In the case of B1929+10, as the pulsar moves through the ISM our HI absorption spectra have 
essentially sampled a one-dimensional cut across the absorbing CNM cloud
(about 50 AU long) and detected at least two TSAS features. This means that a one-dimensional 
covering fraction of TSAS in this direction is at least 40\%, if we assume a 10AU plane-of-sky 
TSAS size. We can now estimate the fraction of the line of sight occupied by TSAS.
We first estimate the line-of-sight size of the parent CNM cloud.
As this cloud has $N(HI)\sim10^{20}$ cm$^{-2}$ (from Table~\ref{t:table_wnm_cnm}), 
and if we assume a CNM 
volume density of 100 cm$^{-3}$, this results in a length of about $7\times 10^4$ AU.
Assuming a cylindrical TSAS with a 10AU plane-of-sky size and a geometrical elongation $G=10$,
 results in the volume filling factor of $<0.2$\% in the direction of B1929+10. The volume filling factor in the 
case of other pulsars in our experiment is zero.
We note that our sensitivity for B0823+26 and B1133+16 is equally good as for
B1929+10, and all three pulsars
probe interstellar environments with a similar optical depth
(peak of $\sim0.2$, see Figures 1 and 2) and a similar line-of-sight length.
In a  similarly sensitive experiment, \cite{Minter05} had 18 observing epochs for one pulsar, about 
150 trials, and {\em{no}} detections.

The observationally inferred volume filling factor of TSAS in the direction of our pulsars (0 or $<0.2$\%)
 is lower than the model prediction ($\sim3$\%).
More importantly, we observe significant inhomogeneity in the galactic distribution of TSAS:
no detections toward four pulsars, and a high detection rate toward B1929+10.
For the Heiles (1997) model to remain viable, our observations require
inclusion of the possibility of different TSAS properties (specifically $L_{\perp}$, $G$ and $\aleph$) along
different lines of sight, and/or additional local physical processes which can modify the
TSAS production rate.
For example, a highly elongated TSAS with $L_{\perp}$ a million times smaller 
than the size of its parent CNM cloud would theoretically result in a volume filling
factor close to zero. While in principle this scenario could work, we are
still left with the puzzle of what physical processes can produce
such tiny clouds and at the same time greatly vary cloud properties across the ISM.

\begin{table*} \setlength{\tabcolsep}{0.02in} \caption{Observed and derived properties of 
TSAS features with $\ga2.5-\sigma$ in the direction of B1929+10.}
\begin{center}
\label{t:tsas_sig_features}
\begin{tabular}{ccccccc}
\noalign{\smallskip} \hline \hline \noalign{\smallskip}
$\Delta t$   &  $l$  & $| \Delta {\rm EW} |$ & Signif. &
$N({\rm HI})_{\rm TSAS}^{a}$ & $n({\rm HI})^{b}$ & $P/k$\\
(yr)      &   (AU)               & (\kms)  & ($\sigma$) & (10$^{19}$ cm$^{-2}$)  & (10$^{4}$
cm$^{-3}$)& (10$^{6}$ cm$^{-3}$ K)\\
\hline
0.16 & 6  &0.046 & 2.8 & 1.4 & 16 & 27 \\
0.33 & 12  &0.083 & 3.3 & 2.6 & 14 & 24 \\
0.75 & 28  &0.040 & 2.5 & 1.2 & 3 & 5 \\
1.24 & 46  &0.089 & 3.5 & 2.8 & 4 & 7 \\
\noalign{\smallskip} \hline \noalign{\smallskip}
\end{tabular}
\end{center}
$^{a}$ We have assumed $T_s=170$ K based on Table~\ref{t:table_wnm_cnm}.\\
$^{b}$ Calculated assuming a spherical geometry.\\
\end{table*}

\begin{table*} \caption{Upper limits on observed TSAS properties from non-detections.}
\begin{center}
\label{t:tsas_upper}
\begin{tabular}{cccccc}
\noalign{\smallskip} \hline \hline \noalign{\smallskip}
PSR &  $l$  & $| \Delta {\rm EW} |$ & $N({\rm HI})_{\rm TSAS}^{a}$ & $n({\rm HI})^{b}$ & $P/k$\\
 &   (AU)               & (\kms)  & (10$^{18}$ cm$^{-2}$)  & (10$^{4}$
cm$^{-3}$)& (10$^{6}$ cm$^{-3}$ K)\\
\hline
B0823+26 & 5-50  &0.05 & $<6$ & $<7$ & $<4$ \\
B1133+16 & 20-170  &0.06 & $<4$ & $<2$ & $<1$ \\
B1737+13 & 20-180  &0.35 & $<19$ & $<7$ & $<4$ \\
B2016+28 & 1-10  &0.45 & $<41$ & $<270$ & $<140$ \\
\noalign{\smallskip} \hline \noalign{\smallskip}
\end{tabular}
\end{center}
$^{a}$ We have assumed $T_s=50$ K.\\
$^{b}$ Calculated assuming a spherical geometry and the largest probed size.\\
\end{table*}

\subsection{The shocked CNM model \citep{Hennebelle07a}}

Several recent numerical simulations of the small-scale structure in the ISM have addressed the 
TSAS phenomenon.
In particular, \cite{Hennebelle07a} performed relatively high resolution numerical simulations of 
colliding WNM flows and the formation of CNM clouds. A collision of incoming WNM streams creates 
a thermally unstable region of higher density and pressure but lower temperature, which further fragments into cold
structures. The thermally unstable gas has filamentary morphology and its fragmentation
into cold clouds is promoted and controlled by turbulence.
While typical CNM clouds are formed in these simulations within $\sim1$ Myr,
transient shocked CNM regions, produced by supersonic collisions between CNM fragments, are 
also found, with properties similar to those of TSAS:
$n\sim10^{3-4}$ cm$^{-3}$,  $T\sim30$ K,  $P/k\sim10^{5}$ K cm$^{-3}$, and a slightly larger size of 
$\sim400$ AU. \cite{Hennebelle07b} found that 10\% of lines of sight cross dense CNM gas with 
$n>10^{3}$ cm$^{-3}$, and
only 0.1\% of lines of sight cross gas denser than $n>10^{4}$ cm$^{-3}$, suggesting that 
extremely dense TSAS should rarely be detected.

While this was a 2D simulation of a coherent region only $20\times20$ pc large, basic 
observed properties of the CNM/WNM were surprisingly well captured. For example,
the CNM column density peak, and the HI column density distribution function appear 
close to observed properties based on the Heiles \& Troland (2003) survey.
Considering that the simulation has a resolution of 400 AU (cruder than the scales 
generally sampled in pulsar and interferometer experiments) and probes regions only 20 pc across
(much smaller than our lines of sight), direct comparison with our observations
is not straightforward.
Nevertheless, the predicted properties of the shocked CNM clouds, especially
cloud abundance, are encouraging.
Considering that the CNM/TSAS production in converging flows is controlled by the level of turbulence 
in colliding flows, local turbulent enhancements  may be able to provide the necessary tuning and therefore explain
the inhomogeneity in the spatial distribution of TSAS observed in the direction of B1929+10.

\begin{figure*}
\epsscale{1.2}
\plotone{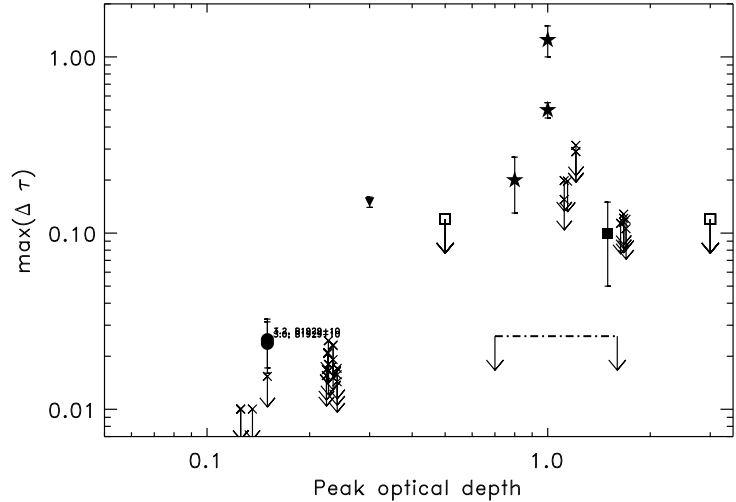}
\caption{\label{f:us_desh1}  Variations in  HI optical depth (and upper limits)
as a function of peak optical depth along the line of sight. Circles (detections)
and x's with arrows (upper limits) are from this study. Filled squares (detections) and 
open squares with arrows (non-detections) are from Johnston et al. (2003).
The horizontal dot-dashed line anchored on either end by arrows represents non-detections over  an almost
continuous sampling of PSR 0329+54 by Minter et al. (2005).
The triangle at a peak optical depth of $\sim0.3$ shows the 2.6-$\sigma$ detection 
in the direction of PSR B0301+19 by
\cite{Weisberg08}. Interferometric detections are shown as stars:
3C138 at a peak optical depth of $\sim0.8$ (Brogan et al. 2005);
3C161, the lower star at a peak optical depth of $\sim1.0 $ \citep{Goss08};
and 3C147,   the upper star at a peak optical depth of $\sim1.0$  (Lazio et al. 2009).}
\end{figure*}

\begin{figure*}
\epsscale{1.2}
\plotone{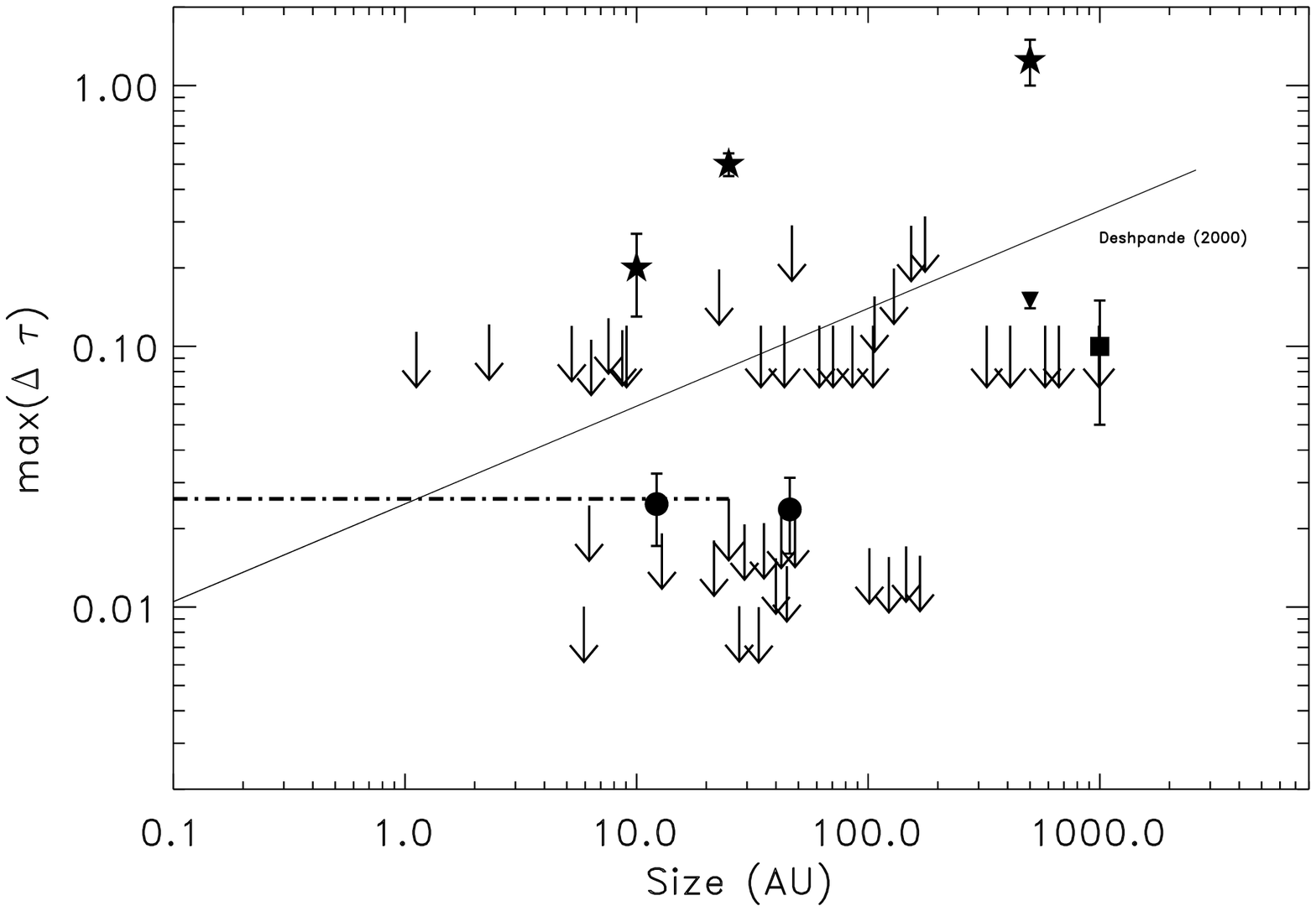}
\caption{\label{f:us_desh2}
Variations in HI optical depth (and upper limits) as a function of spatial
scale. Symbols are the same as in the previous figure.
The interferometric detections (stars:
3C138 at 25 AU (Brogan et al. 2005);
3C161 at 500 AU \citep{Goss08}; and
3C147 at 10 AU (Lazio et al. 2009).
The solid line shows the level of
fluctuations in $\Delta \tau$ as predicted by Deshpande (2000).
The dot-dashed line represents an almost
continuous sampling of PSR 0329+54 by Minter et al. (2005) with no detections.
}
\end{figure*}

\subsection{The cut-off scale of the CNM turbulence?}

In an attempt to compare various observations, we have investigated HI TSAS detections and upper limits from
all published and recent pulsar and interferometric experiments as a
function of their spatial scale.  The results are
summarized in Figures~\ref{f:us_desh1} and ~\ref{f:us_desh2}.

In Figure~\ref{f:us_desh1} we first show the maximum variation in HI optical depth ($\Delta \tau$) as
a function of peak optical depth along the line of sight ($\tau_{\rm peak}$). This plot 
separates all sources cleanly into two groups:
``low optical depth'' with $\tau_{\rm peak}<0.5$ (B1929+10, B0823+16, B1133+16, 
and B0301+19), and ``high optical depth'' with $\tau_{\rm peak}>0.5$
(3 interferometric detections, pulsars B1737+13 B2016+28 from
this study, highly sampled pulsar B0329+54 from Minter et al. 2005, and B0736-40, 
B1451-68, B1557-50 from Johnston et al. 2003). The only exception in this figure is the 2.6-$\sigma$ 
detection by Weisberg et al. (2008)
in the case of B0301+19 which probes a low optical depth CNM but has $\Delta \tau \sim0.1$.
This plot suggests that regions with high optical depth (and therefore high CNM column density) 
have higher level of $\Delta \tau$ variations than regions
with low CNM column density. In other words, this implies that the the level of $\tau$ 
fluctuations tracing TSAS is intimately
associated with the larger scale structure of the parental CNM.

Figure~\ref{f:us_desh2} shows  the detected level of
optical depth variations, or upper limits, as a function of TSAS spatial scale $l$.
No obvious correlation between the two quantities is noticeable. The Kendall's tau correlation test
for censored data \citep{Isobe86} gives a probability of 62\% that the correlation is not present.
The low and high $\tau_{\rm peak}$ groups of sources nicely separate in this plot around
$\Delta \tau=0.03$: the high-$\tau_{\rm peak}$ sources are above this line, while all 
low-$\tau_{\rm peak}$ sources except B0301 are below this line.
We investigated whether a correlation between TSAS scale $l$ and $\Delta \tau$ exists for the two
groups separately. For both groups the Kendall's tau correlation test
for censored data gives a probability of $\sim50$\% that the correlation is not present.

We therefore conclude that the detected level of optical depth variations, as well
as upper limits, do not show evidence for a correlation between $l$ and $\Delta \tau$.
Such a correlation, with $\Delta \tau \propto l^{(\alpha-2)/2}$,
is expected in the Deshpande (2000) power law model, and would represent
the tail-end of the turbulent spectrum of the HI optical depth on larger scales (0.02 to 4 pc, Deshpande et al. 2000).
We overplot this predicted level of variations
in Figure~\ref{f:us_desh2} as a sloping line.

Our compilation of recent interferometric and pulsar observations of TSAS does not
provide evidence for the power-law turbulent spectrum of $\Delta \tau$ on scales
$<1$ to 1000 AU. While various sources probe separate lines of sight, we note that
B0823+26, B1133+16 and B1929+10 all probe the local ISM (within 400 pc), with similar
peak optical depth, and with very similar sensitivity. This essentially provides
an almost uniformly sampled region on this graph from $\sim10$ to 200 AU.
Similar arguments apply to the higher optical depth portion of this diagram.
We are left with the conclusion that TSAS is a sporadic phenomenon.

It is interesting to note that the smallest spatial scale of TSAS detections in both 
interferometric and B1929+10 studies is $\sim10$ AU.
Two pulsars (B2016+28 and B0329+54) have probed scales $<10$ AU, however TSAS 
was not detected. The spatial scales of $\sim10$ AU for TSAS in the CNM are especially interesting, since
in the case of hydrodynamic turbulence
the cut-off scale of CNM turbulence corresponds to the
viscous damping scale of $\sim20$ AU \citep{Zweibel05}.
However, in the case of magnetohydrodynamic (MHD) turbulence, the turbulent spectrum can
cascade below the viscous cut-off scale down to the scale set by magnetic diffusion
\citep{Cho02}. In fact, magnetic and kinetic spectra can be greatly decoupled, with the
magnetic spectrum being much shallower than the Kolmogorov spectrum.

\subsection{What's going on with TSAS?}

Our pulsar experiment shows only a few significant TSAS detections, all in the direction
of one pulsar --- B1929+10. This result strongly
suggests inhomogeneity in the galactic distribution of TSAS.
We have examined two proposed origins of TSAS --- (i) discrete HI blobs elongated along the
line of sight;   and (ii) fluctuations at the tail-end of the turbulent CNM spectrum; but our data do not
provide support for either of these models.  The principal problem is that we do not find 
these features to be homogeneously distributed. Meanwhile, recent numerical simulations 
do generate CNM clouds with a wide range of physical
properties, including rare examples of very dense and over-pressured TSAS-like  clouds; but
the direct comparison between observations and simulated data is not straightforward
since simulations probe much smaller ISM regions on scales somewhat larger than these experiments.

The most striking result from our examination of specific lines of sight is a possible correlation 
between the B1929+10 fluctuations and
NaI interstellar clouds   inside the Local Cavity or at the LB wall.
As about one-half of B1929+10's line of sight is associated with the LB, and a significant fraction 
of this length appears to be traversing the LB wall, we suggest that the observed HI absorption 
fluctuations are most likely sampling small-scale structure in the LB. Numerous observations have 
indicated that the LB wall is highly inhomogeneous.
Many interstellar cloudlets found inside the Local Cavity are thought to be caused by 
hydrodynamic instabilities fragmenting
the LB wall. Through various studies, it has been shown that these cloudlets have velocities 
of up to 25 \kms~relative to the Sun \citep{Muller06}. This translates to $\sim5$ AU yr$^{-1}$. Coupled 
with the transverse speed of B1929+10 of 37 AU yr$^{-1}$, this suggests that our observed TSAS could 
result from the pulsar's
line of sight crossing very local cloudlets.

To investigate the origin of interstellar clouds inside the Local Cavity,
we look to 2D hydrodynamical simulations of the WNM compression by supernova explosions 
and subsequent fragmentation into small clouds \citep{Koyama02}. They showed that within 0.3 
Myr a thermally collapsing layer is formed behind the (supernova) shock front
where cooling dominates heating and temperature monotonically decreases.
As the gas cools it becomes thermally unstable, with a temperature in the range of
300-6000 K. Within $\sim1$ Myr this layer starts to fragment into many small ($< 10^{-2}$ pc) cloudlets.
Finally, these cloudlets can even reach
thermal equilibrium at a temperature of 20 K and density of 2000 cm$^{-3}$. As the ISM is
 frequently compressed by supernova
explosions, fragmentation of the shock-compressed medium can provide a mechanism for the
 formation of many small cloudlets.

In the \cite{Koyama02} simulation, the cloudlets become embedded in the hot
high-pressure gas as the shocked layer fragments, leading to a situation similar to what is observed in the LB.
The exact temperature inside the Local Cavity has
been uncertain (see \cite{Welsh10b} for references),
 however it is clear that the Cavity is occupied
by a warm/hot ionized medium.
The cold cloudlets will eventually evaporate in the surrounding warm/hot medium.
This process has been studied by many authors.
The mass-loss rate in the case of classic evaporation \citep{McKee77b} 
is slightly lower when cooling is included \citep{Nagashima06}.
The mass-loss rate of \cite{Slavin07} who considered both flux saturation effects and
radiative cooling is about 100 times lower than the classic evaporation rate.
We continue by assuming the lowest suggested mass-loss rate, which
will have the gentlest effect on the cold clouds.

The evaporation timescale depends greatly on the cloud size and the temperature of the 
surrounding medium. For a cloud size of 30 AU, $\tau_{\rm evap}=10^2$ - $10^{3}$ yr, 
depending upon whether the cloud is surrounded by the hot ionized medium (HIM) or WNM.
For a larger cloud with a size of $10^4$ AU, $\tau_{\rm evap}=10^{5-6}$ (HIM) or 10$^{6-7}$ (WIM) yr.
Clearly larger clouds survive longer and can even coalesce to build larger structures, even if
surrounded by a hot medium. In about 1 Myr larger clouds could cross a large distance, $\sim20$ pc, 
and may even escape outside of the LB into the ISM through various chimneys and tunnels resulting 
from stellar winds and supernovae, and frequently
observed in expanding shells.
This scenario may be able to explain the existence of HI clouds on scales of thousands of AU
or more in the ISM, as observed recently by \cite{Stanimirovic05}. On the contrary, very small clouds 
will evaporate quickly and therefore may not be common
in the ISM. Their short inferred lifetimes indicate that they would usually be 
found close to their formation sites.

\section{Summary}
\label{s:conclusions}

We have used the Arecibo telescope to obtain multi-epoch HI absorption spectra
in the direction of six bright pulsars, with the goal of searching for structure on AU scales in the CNM. 
Our target sources were observed previously by Frail et al. (1994),
who found pervasive tiny-scale HI structure.
We used an advanced spectrometer capable of more accurate measurements in the
face of wildly fluctuating pulsar signals, and carefully analyzed noise statistics.
To search for variability of HI absorption profiles, which signifies inhomogeneities
on AU scales in the CNM, we have searched for time-variability of the  absorption line
equivalent widths and of the absorption spectra themselves.
While we had excellent sensitivity, significant variations
in both EW and absorption spectra have been found only in the case of
one pulsar: B1929+10. These variations imply TSAS on spatial scales
$l=\frac{D_{\rm HI}}{350~{\rm pc}} \times (6, 12, 28, 46)$ AU (where $D_{\rm HI}$ is the distance to the
TSAS feature), with
an inferred volume density  of $>10^4$ cm$^{-3}$ and  thermal pressure of
$>10^6$ cm$^{-3}$ K.

Our study clearly shows the inhomogeneity in the
galactic distribution of TSAS: the detection rate toward four of our pulsars is zero, while the detection rate 
toward B1929+10 is high.
Such inhomogeneity is hard to explain with the model
of discrete elongated disks or cylinders being a {\it{general}}
property of CNM clouds (Heiles 1997); instead
large spatial variations of TSAS physical properties or abundances must be invoked.
While the Heiles (1997) model  can help in bringing down the volume density and pressure of TSAS, 
the puzzles regarding its distribution and origin still remain.

Our examination of all pulsar and interferometric TSAS detections and upper limits
does not show a correlation between the level of optical depth fluctuations and the TSAS spatial scale,
as would be expected if the turbulent spectrum on
much larger scales is extrapolated to AU-scales.
The detections and non-detections probe an almost continuous range
of spatial scales from $\sim0.1$ to 1000 AU.
Therefore, the large number of non-detections of TSAS suggests that the CNM clouds 
on scales $10^{-1}$ to $10^{3}$ AU are not a pervasive property of the ISM.
The sporadic TSAS detections on scales of tens of AU may indicate occasional local bursts of 
turbulent energy dissipation,  rather than end-points of the turbulent spectrum.

Another striking result from our study is a possible correlation between B1929+10's TSAS and
 interstellar clouds observed in NaI absorption inside the Local Bubble.
We find evidence that the TSAS is likely to be within 106 pc of the Sun, and is
sampling the small-scale structure of the LB caused by hydrodynamic instabilities fragmenting the LB wall.
We propose that the line of sight of B1929+10 is revealing
this recently formed small-scale structure. Similar bubbles and their walls are found throughout
 the Milky Way,  but the lifetime of a TSAS cloud created from them depends strongly on the 
 its size and the temperature of the surrounding medium.
Larger fragments (size $\sim10^4$ AU) survive longer and can travel large ISM distances, 
becoming a more general ISM property. On the other hand, the smallest clouds (size $\sim10-100$ AU)
evaporate quickly close to their formation site, and are therefore
not very commonly observed in the ISM. Additional processes, such as stellar
mass-loss and collisions
of interstellar clouds/filaments, probably contribute to
the CNM structure formation on somewhat larger (sub-pc) scales.

\begin{acknowledgements}
We are grateful to Caltech's Center for Advance Computation and Research
for the use of their facilities for data storage and processing.
We express our thanks to Stuart Anderson, Rick Jenet, and Kathryn Devine for help
with observations and data processing. We acknowledge the use of the ATNF Pulsar Catalog.
We thank Carl Heiles, J.-P. Macquart, Miller Goss, Jay Gallagher, and Shu-ichiro Inutsuka 
for stimulating discussions.
We would also like to thank the anonymous referee, whose comments 
improved greatly the clarity of the paper.
SS acknowledges support from NSF grants  AST-0097417 and AST-9981308, 
and the Research Corporation. She also thanks ASTRON and the Helena Kluyver visiting 
program for their hospitality during 
the final stage of the manuscript preparation.
JMW, ZP, KT, and JTG were supported by NSF Grants  AST-0406832 and 
AST-0807556. \end{acknowledgements}

\bibliographystyle{/d/leffe/sstanimi/rattler/TexStyles/apj}

\label{lastpage}
\end{document}